\documentclass[10pt,superscriptaddress,nofootinbib]{revtex4-2} 

\usepackage{ragged2e}

\usepackage{multirow}

\usepackage{amsmath}  
\usepackage{mathtools}
\usepackage{amsfonts} 
\usepackage{graphicx} 
\usepackage{xcolor}

\usepackage{array}
\usepackage{makecell}

\usepackage{footnote}

\usepackage[justification=justified,labelfont=bf,format=plain]{caption}
\DeclareCaptionFormat{myformat}{\fontsize{10}{10} \selectfont#1#2#3}
\begin{document}


\title{Adaptive High-Level Tight Control of Prostate Cancer: \\ A Path from Terminal Disease to Chronic Condition}

\author{Trung V. Phan}
\email{tphan@natsci.claremont.edu}
\affiliation{Department of Natural Sciences, Scripps and Pitzer Colleges, Claremont Colleges Consortium, Claremont, California 91711, USA}

\author{Shengkai Li}
\affiliation{Department of Physics, \\ Princeton University, Princeton, NJ 08544, USA}

\author{Luciana Sarabia}
\affiliation{Scripps College, Claremont, CA 91711, USA}

\author{Caroline N. Cappetto}
\affiliation{Scripps College, Claremont, CA 91711, USA}

\author{Benjamin Howe}
\affiliation{Department of Physics, \\ Princeton University, Princeton, NJ 08544, USA}

\author{Sarah R. Amend}
\affiliation{Cancer Ecology Center, The Brady Urological Institute, \\ Johns Hopkins School of Medicine, Baltimore, MD 21287, USA}

\author{Kenneth J. Pienta}
\affiliation{Cancer Ecology Center, The Brady Urological Institute, \\ Johns Hopkins School of Medicine, Baltimore, MD 21287, USA}

\author{Joel S. Brown}
\affiliation{Departments of Radiology and Integrated Mathematical Oncology, \\ Moffitt Cancer Center,  Tampa, FL 33612, USA}

\author{Robert A. Gatenby}
\affiliation{Departments of Radiology and Integrated Mathematical Oncology, \\ Moffitt Cancer Center,  Tampa, FL 33612, USA}

\author{Constantine Frangakis}
\affiliation{Department of Medicine, Johns Hopkins University, \\ 
Baltimore, MD 21218, USA}
\affiliation{Department of Biostatistics, Bloomberg School of Public
Health, \\ Johns Hopkins University, 
Baltimore, MD 21218, USA}

\author{Robert H. Austin}
\affiliation{Department of Physics, \\ Princeton University, Princeton, NJ 08544, USA}

\author{Ioannis G. Keverkidis}
\affiliation{Department of Chemical and Biomolecular Engineering, \\ John Hopkins University, Baltimore, MD 21218, USA}

\begin{abstract}
Metastatic prostate cancer is one of the leading causes of cancer-related morbidity and mortality worldwide. It is characterized by a high mortality rate and a poor prognosis. In this work, we explore how a clinical oncologist can apply a Stackelberg game-theoretic framework to prolong metastatic prostate cancer survival, or even make it chronic in duration. We utilize a Bayesian optimization approach to identify the optimal adaptive chemotherapeutic treatment policy for a single drug (Abiraterone) to maximize the time before the patient begins to show symptoms. We show that, with precise adaptive optimization of drug delivery, it is possible to significantly prolong the cancer suppression period, potentially converting metastatic prostate cancer from a terminal disease to a chronic disease for most patients, as supported by clinical and analytical evidence. We suggest that clinicians might explore the possibility of implementing a high-level tight control (HLTC) treatment, in which the trigger signals (i.e. biomarker levels) for drug administration and cessation are both high and close together, typically {\color{black}yields} the best outcomes, as demonstrated through both {\color{black}computation} and theoretical analysis. This simple insight could serve as a valuable guide for improving current adaptive chemotherapy treatments in other hormone-sensitive cancers.
\end{abstract}

\date{\today}

\maketitle 

\section{Introduction}

The analysis that follows of our proposed adaptive high-level tight control (HLTC) treatment of resistant {\color{black}prostate} cancer predicts improved {\color{black}outcomes} compared to conventional therapies, either by maximizing the total drug response time before progression or, most importantly, transforming the disease into a chronic condition for a wide range of patient conditions. This prediction could serve as a guide to improve current adaptive chemotherapy treatments. We suggest that clinicians might explore the possibility of implementing HLTC, i.e. waiting for the disease to advance further rather than initiating treatment too early. This kind of ``watchful waiting'' approach before treatment is not uncommon in practice, at least historically \cite{chapple2002watchful}, but we take it a large step further. Our simulations indicate that, with the approach developed in the following, 19 of the 32 patients we analyzed could potentially have become chronic prostate cancer survivors.  In what follows, we present a detailed analysis of how it is possible to make the disease chronic.
 
 The central approach of this paper {\color{black}focuses} on competition between populations within confined ecological niches and {\color{black}on how clinicians can regulate these populations by responding to changes in their relative abundances}. Cancer ecosystems are characterized by their carrying capacities, that is, the maximum number of cancer cells that can be supported by the environmental resources available in that niche is finite and computable.

Within the prostate gland, prostate tissue cells are sensitive to male androgens {\color{black}(i.e. male sex hormones such as testosterone that regulate cell growth and death in the tissue)}, serving as key regulators of cell death and growth in the tissue \cite{androgens}. In a healthy state, the growth and death rates of prostate tissue cells are controlled by androgen levels, ensuring glandular balance. In contrast, prostate cancer cells disregard these androgen controls and continue to grow.
Hormone deprivation therapy{\color{black}, a treatment strategy that lowers or blocks androgen signaling,} targets androgen-dependent prostate cells by cutting off the supply of androgen growth hormones to limit their growth \cite{deprivation}. However, prostate cancer cells can acquire resistance to this treatment, enabling them to proliferate even in the absence of these hormones. Such resistant cells may (1) preexist within the prostate tumor prior to initial therapy due to genetic or epigenetic diversity \cite{turke2010preexistence,bhang2015studying,sottoriva2015big}, or (2) emerge as {\em de novo} resistant cells through natural selection \cite{fitzgerald2017stress,hoffmann2000environmental,hata2016tumor,ramirez2016diverse}. 

The two distinct cell types, which are sensitive (+) or resistant (-) to androgens, can compete within a shared environment, creating a scenario similar to a resource game \cite{resource_game_theory}. Although the overall carrying capacity is dictated by the maximum number of cells that can occupy a particular environment, the proportion of androgen-sensitive and -resistant cells may vary on the basis of their competitive interactions.

Testosterone serves as the primary male androgen. Androgens such as testosterone attach to androgen receptors found on the surfaces of prostate cells. Through a complex signaling pathway, the occupation of these receptors stimulates the proliferation and apoptosis of androgen-sensitive cells. These cells, whose growth and demise are regulated by testosterone, are known as T$^+$ cells due to their dependence on bound androgen receptors for growth and survival \cite{androgen_sensitive}.

Over time, prostate tissue can develop or already contain cells ({\color{black} both sensitive and resistant}) with an increased number of androgen receptors, increasing their sensitivity to testosterone and allowing them to proliferate with lower testosterone levels. Because these cells can thrive even when testosterone is reduced, they are labeled ``resistant'' to the agonist at pituitary receptors drug leuprorelin {\color{black}(in which an agonist is a molecule that binds to and activates a receptor, and leuprorelin is a hormone-deprivation drug that lowers testosterone production}) \cite{leuprorelin_action}, although their resistance is indirectly related to the drug itself \cite{leuprolide_resistance}. This resistance can progress, leading to prostate cancer cells that can grow entirely without testosterone, known as T$^-$ cells. These cells are not effectively managed by standard androgen deprivation therapy, which means that even when administered leuprorelin levels are elevated, T$^-$ prostate cancer cells continue to proliferate.

 Prostate specific antigen (PSA){\color{black}, a blood biomarker commonly used to monitor prostate cancer burden,} is generated by both normal prostate cells and cancerous ones. An increase in PSA levels in the bloodstream, which suggests unusual proliferation of prostate cells, prompts the initial therapeutic approach of administering leuprorelin. Leuprorelin functions as an agonist in the pituitary gland's gonadotropin-releasing hormone (GnRH) receptors, ultimately leading to a reduction in testosterone production by testicular cells, thus inhibiting growth and leading to apoptosis of T$^+$ prostate cells. Individuals undergoing prolonged leuprorelin treatment are described as "chemically castrated." However, this state is completely reversible upon cessation of leuprorelin, in contrast to the permanent nature of surgical castration.
 
One reason patients treated with chemical castration using leuprorelin may show continuous PSA increases in metastatic castration-resistant prostate cancer (mCRPC){\color{black}, an advanced form of prostate cancer that continues to progress despite testosterone-suppressing treatment,} is the adrenal glands' production of testosterone \cite{adrenal_production}. To combat this, abiraterone acetate{\color{black}, a drug that inhibits androgen synthesis,} is given to inhibit androgen synthesis in the adrenal glands, further lowering testosterone levels than leuprorelin alone can achieve. Abiraterone works by inhibiting the CYP17A1 enzyme, which is crucial for the production of androgen (male hormone) in the testes, adrenal glands, and prostate cancer tissues \cite{de2011abiraterone}. By blocking this enzyme, the drug drastically decreases androgen steroid hormones, notably testosterone, which significantly influence the growth potential of these susceptible cells.
 
 The patients we are concerned with here are all patients with mCRPC who still have populations of T$^+$,  T$^{+/-}$ and T$^-$. 
 The task in this paper is to optimize the suppression of T$^-$ cell growth through competition with T$^+$ and T$^{+/-}$ cells for mCRPC patients. 

Cancer adaptive chemotherapy, as outlined by Gatenby \cite{gatenby2009adaptive}, can be interpreted through the framework of Stackelberg sequential game theory \cite{stackelberg}. In this approach, an oncologist exploits the competitive struggle for resources among cancerous cells with the goal of fostering the growth of drug-sensitive subclones. These subclones, in turn, help curb the expansion of drug-resistant populations \cite{west2020towards,west2019multidrug}. 

The key idea of adaptive chemotherapy is primarily the interplay between diverse cell groups, which compete for (1) a finite cell carrying capacity and (2) the degree of rivalry among cells, even when the carrying capacity has not yet been reached. In both cases, whether nearing carrying capacity or in the presence of interpopulation competition, a more strategically dominant cell subpopulation can inhibit the growth of a strategically minor subpopulation. This strategic suppression may also impede the expansion of a fast-replicating, drug-resistant variant. In scenarios characterized by competitive limitation of growth rates near carrying capacity, adaptive chemotherapy can postpone the emergence of resistant cell groups \cite{hansen-read}. However, as we will illustrate, when competition coefficients are sufficiently in favor of sensitive cells, it is possible to continuously suppress resistant cell populations, even when they are far from reaching their carrying capacity. 

In adaptive chemotherapy, the clinician dictates the treatment strategy through dose scheduling. Ideally, both dose intensity and duration could be adjusted. However, if dose intensity remains fixed, then the PSA \textit{on-level} and \textit{off-level} serve as two modifiable trigger parameters to modify drug dosage. Currently, in clinical adaptive treatments, these adjustable parameters have not been optimized for maximum drug efficiency or response duration \cite{gatenby2009adaptive} and are instead set at an arbitrary 50\% of the baseline (initial) PSA level of the incoming patient \cite{zhang2022evolution}.

As Hansen and Read have noted \cite{hansen-read}, it is better to select the PSA \textit{on-level} for administering drugs to be above the initial baseline measurement at the beginning of treatment. This choice should ensure that the count of drug-sensitive cancer cells, T$^+$, approaches the total carrying capacity, while T$^-$ remains significantly below it. In a straightforward model with two populations—where the first group, comprising T$^+$ and T$^{+/-}$, is sensitive to drugs, and the second group, consisting only of T$^-$, is resistant—the PSA \textit{off-level} should be nearly equal to the \textit{on-level}. Both levels need to be near the carrying capacity to keep drug-sensitive cells close to their maximum and to limit the proliferation of drug-resistant cells.

These two concepts, (1) positioning the PSA trigger \textit{on} level to maintain the total cell number near the total body carrying capacity and (2) setting the differential between \textit{on} and \textit{off} as small as possible are not new \cite{hansen-read}. What we have further explored and propose here are:
\begin{itemize}
    \item A systematic statistical approach to determine, from clinical data,  important parameters describing cancer progression, including personalized growth rates and competition coefficients, as well as personalized parameters such as the initial number of cancer cells relative to carrying capacity and drug effectiveness, in the fundamental game theory model described in Section \ref{Stackelberg_game_theory}. The details are provided in Section \ref{Extraction}.
    \item The possibility, in favorable patient cases, of extending the period of cancer suppression below a threshold or even making prostate cancer chronic rather than progressive, albeit with continued therapy throughout the patient’s lifetime. We show how these outcomes can be achieved via a \textit{high level tight control} adaptive chemotherapy in Section \ref{Optimization} using Bayesian optimization. We provide analytical justifications and estimates for these findings, see Appendix \ref{analytic}.
\end{itemize}

\section{A Simplified Stackelberg Game Theory Model of\\
Adaptive Therapy for Prostate Cancer \label{Stackelberg_game_theory}}

 Consider an ensemble of $N_k(t)$ cancer cell subpopulations (indexed by $k$);  the total number of cancer cells is given by $N(t) = \sum_k N_k(t)$, and the PSA level is assumed to obey the following ordinary differential equation (ODE):
\begin{equation}
\frac{d}{dt} \text{PSA}(t) = \lambda N(t) - \Xi \ \text{PSA}(t) \ ,
\label{PSA_change}
\end{equation}
where the PSA decay rate is $\Xi \sim 3.5\times 10^{-1}$/day (corresponding to a characteristic decay timescale of about a few days \cite{oesterling1993effect}) and $\lambda$ is the PSA production rate per cell, assumed uniform across sub-populations. It is crucial to note that not all cancer cells produce PSA in the same way or at consistent levels over time; here we are relying on a rather simplistic assumption for such a complex disease to guide our estimation for parameters that describe cancer dynamics. Since the rate $\Xi$ is significantly larger than the cancer cell growth rates (which ranges from $2\times 10^{-3}$ to $2\times 10^{-2}$/day, as estimated in \cite{zhang2022evolution}), we can further adopt the additional quasi-steady state approximation (QSA):
\begin{equation}
\text{PSA}(t) \approx \left(\lambda /\Xi\right) N(t) \ ,
\label{PSA_change_QSA}
\end{equation}
 meaning that the system is considered highly overdamped.

The oncologist (the ``game leader''), begins somewhere in the game at an unknown point, presented with a patient showing elevated PSA levels, that is we do NOT know the carrying capacity of the patient's tumor. Let us consider a simplified version for the basic case of \textit{only two} prostate cancer cell sub-populations under the influence of a \textit{single} drug: (1) a drug-sensitive population $N_S(t)$, and (2) a drug-resistant population $N_R(t)$. In this model, we assume no treatment- or time-dependence to their phenotypes. That is, there is no {\em de novo} resistance evolution: no phenotype switching or {\color{black}mutagenic} response in the cancer cells to hormone deprivation. Both cell phenotypes share a common cell carrying capacity $K$ in the absence of treatment. The cancer dynamics is then described by the following two simple coupled ODEs \cite{zhang2022evolution}:
\begin{equation}
\begin{split}
\frac{d}{dt} N_S(t) = r_S \left[ 1-\bigg(\frac{N_S(t) + N_R(t)}{K\left[ 1-\gamma \Lambda(t)\right]} \bigg)\right] N_S(t) & \ ,
\\
\frac{d}{dt} N_R(t) = r_R \left[ 1-\bigg(\frac{N_R(t) + \alpha_{RS} N_S(t)}{K} \bigg) \right] N_R(t) & \ ,
\end{split}
\label{2ODEs_model}
\end{equation}
in which $r_S, r_R>0$ are maximum growth rates of sensitive and resistant cells;
the cells compete for a resource (testosterone) with the competition coefficient $\alpha_{RS} \geq 0$. This coefficient {\color{black}represents} how the two cell types interact and affect each other indirectly through their distinct metabolisms and influence on the tumor microenvironment \cite{tumor-micro}. In Eq. \eqref{2ODEs_model}, the effect of the drug is to reduce the carrying capacity for sensitive cells by a factor of $\gamma \Lambda(t)$, where $\Lambda(t)$ denotes the time-dependent level of the drug administered to the patient. The value $\gamma$ represents the drug effectiveness, $0\leq \gamma \leq 1$. We take $\Lambda(t)$ to be \textit{binary}: $\Lambda(t)=0$ during periods of no drug administration, and $\Lambda(t)=1$ during periods of (maximum) dosage. The use of $\gamma \Lambda(t)$ may appear \textit{ad-hoc}, but it is predicated {\color{black}on} the presence of Abiraterone, which reduces the carrying capacity by a factor of $(1-\gamma)$ for only drug-sensitive T$^+$ populations, but stays unchanged for resistant T$^-$ populations. 

Eq. \eqref{PSA_change_QSA} and Eq. \eqref{2ODEs_model} contain six parameters, i.e. $\{ r_S, r_R, \alpha_{RS}, \gamma, K, \lambda \}$. Presumably all these parameters are \textit{patient-specific}. In addition, there are at least two more unknowns, related to the \textit{initial} cancer population sizes $\{ x_S(0),x_R(0) \}$ for each patient. It is important to note that at $t=0$ for each patient, they have already received other treatments (e.g., chemical castration) in preparation for the study. The modeler must determine parameter values \textit{of each patient}. In previous work by Zhang et al. \cite{zhang_2017}, finding unique patient-specific parameters was not done systematically. Simulations were carried out assuming all patients entered the trial {\color{black}with tumor burden near one-half of the carrying capacity}, guaranteeing that adaptive therapy would have a profound effect. This assumption of being close to carrying capacity in the simulations was carried forward into the clinical trials, where PSA levels were set to a normalized value of 1 for all patients to be near the carrying capacity. There was no attempt to actually determine carrying capacity from a dose-response experiment. In subsequent work by Zhang et al. \cite{zhang2022evolution}, the parameters have been determined systematically but under an arbitrary choice for drug effectiveness and an assumption that the cancer growth rates are identical across patients, which is not supported by the data. Furthermore, the trigger points for on/off for abiraterone therapy were arbitrarily set (the PSA off-level is half that of the PSA on-level), with no attempt to optimize these values.

 Hansen and Read et al. \cite{hansen-read} realized that the Zhang et al. \cite{zhang2022evolution} lacked optimization.  They pointed out that actually setting the off trigger close to the carrying capacity, and setting the on {\color{black}trigger} as close as possible to the one one actually could significantly improve patient time to progression, which of course makes sense since maintaining the $T^+$ population at as high a value as possible within the confines of this simple model can only help suppress the T$^-$ population. This was an {\color{black}important} advance, but unfortunately Hansen and Read presented no systematic procedure for actually determining carrying capacity numbers from patient data.  As they point out, if one starts far from the carrying capacity and with closely matched populations of  T$^+$ and T$^-$ an adaptive protocol can actually suppress the T$^+$ population and help the T$^-$, a disaster.

 The only way to achieve a true patient specific model of that patient's disease state is through experimentation, i.e. administering brief doses of Abiraterone and measuring PSA levels over time to fit all the parameters to the measured PSA trajectories for each patient. Once parameters are established, the oncologist can use them to project the future of cancer progression for each patient in a Stackelberg game, and can even perform treatment optimization and open-loop planning. To begin, we need to make some assumptions to reduce the number of free parameters, ensuring that the fit remains predictive while avoiding excessive aggregating/simplification that could lead to an inaccurate prediction of cancer progression under treatment.

 \section{A More Detailed Stackelberg Game Theory Model of\\
Adaptive Therapy for Prostate Cancer}

A central problem in achieving a truly personal, patient-specific model for treating mCRPC is the large number of free parameters. 
For a patient with fewer than $\sim 10$ PSA blood level measurements (typically, a patient in \cite{zhang2022evolution} has around $\sim 20$ measurements), the parameters in the ODEs in Eq. \eqref{PSA_change_QSA} and Eq. \eqref{2ODEs_model} cannot be reliably identified when fitting observational data. We therefore now consider constraining these parameters while being careful not to arbitrarily fix too many of them.

For any individual patient, we can re-write the system of ODEs describing the evolution of the observable PSA level and the cancer progression, i.e. Eq. \eqref{PSA_change_QSA} and Eq. \eqref{2ODEs_model}, as follows:
\begin{equation}
\begin{split}
 \text{PSA}(t) \approx \tilde{\lambda} x(t) & \ , 
\\
\frac{d}{dt} x_S(t) = r_S \left[ 1-\bigg( \frac{x_S(t) + x_R(t)}{ 1-\gamma \Lambda(t)} \bigg) \right] x_S(t) & \ ,
\\
\frac{d}{dt} x_R(t) = r_R \left[ 1 - \Big(x_R(t)+\alpha_{RS} x_S(t) \Big)\right] x_R(t) & \ ,
\end{split}
\label{prostate_cancer_small_model}
\end{equation}
in which $x_S(t), x_R(t)$, and $x(t) \equiv x_S(t) + x_R(t)$ are the sensitive, resistant, and total cancer cell populations respectively, scaled by a common carrying capacity $K$:
\begin{equation}
x_S(t)\equiv N_S(t)/K \ , \ x_R(t)\equiv N_R(t)/K \ . 
\end{equation}
The quantity $\tilde{\lambda} \equiv (\lambda/ \Xi)K $ is the ratio of the PSA production rate to its decay rate when the cancer reaches its carrying capacity; this is also equal to the carrying capacity PSA measurement under QSA:
\begin{equation}
\text{PSA}_K \equiv \text{PSA}\Big|_{x=1} = \tilde{\lambda} \ .
\label{PSA_K}
\end{equation}
It is important to note that cancer can cause symptoms, complications, and even become lethal \textit{before} $x(t)$ reaches $1$.


Assuming that the above description is adequate across patients, then if the seven parameters $\{ \tilde{\lambda}, r_S, r_R, \alpha_{RS}, \gamma ,x_S(0), x_R(0) \}$ are known for each and every patient, the PSA trajectories can be fully determined using Eq. \eqref{prostate_cancer_small_model} for any drug treatment schedule $\Lambda(t)$. Note that all of these parameters may differ between patients; however, by assuming some are common, we can still achieve a good fit for all patients and, most important, infer some general behavior of prostate cancer biophysics to guide and optimize treatments. Also, not all of these parameters are of equal importance. As we will discuss in Section \ref{Optimization}, the parameters that have the most influence on patient outcomes, when treatment is optimal, are the \textit{common} ecological competition coefficient $\alpha_{RS}$ and the \textit{patient-specific} biomarker level at carrying capacity $\text{PSA}_K=\tilde{\lambda}$. 

Consider the \textit{relative} kinetics between sensitive and resistant cells to be universal across patients. This means that the growth and competition of sensitive and resistant cells are locally identical across patients, differing only by a timescale. We can thus further divide the seven parameters into two sets:
\begin{itemize}
    \item The two \textit{common parameters} $\{ r_S/r_R, \alpha_{RS}\}$, representing the ratio between growth rates and the ecological competition coefficient.
    \item The five \textit{patient-specific parameters} 
    \begin{equation}
        \{ \tilde{\lambda}, r_R, \gamma, x_S(0), x_R(0) \} \ ,
    \label{5_ps_params}
    \end{equation}
    which are the PSA measurement when the cancer population reaches carrying capacity, the resistant cell growth rate, the drug effectiveness, and the initial populations sizes of the two different cancer cell phenotypes at the beginning of PSA tracking for each individual patient. This is the smallest set of parameters that should not be identical across patients.
\end{itemize}
In contrast to the previous study \cite{zhang2022evolution}, here we acknowledge that cancer growth and patient responses to drug may vary drastically between individuals. These considerations are consistent with crude estimations of growth rates for a subset of patients \cite{zhang2022evolution} and with findings related to Abiraterone responses \cite{fizazi2017abiraterone,hashimoto2019serum}. We also did not fix any parameter values before fitting the model, as \cite{zhang2022evolution} did for $\gamma=90\%$ and $r_S/r_R=1.66$ \cite{zhang2022evolution}; instead, we obtain these values from the fitting procedure. Note that we did explore other options for both \textit{common parameters} and \textit{patient-specific parameters}, but the selection above appears to provide the best fit.\footnote{For example, allowing all parameter to be patient-specific or setting $\alpha_{SR}=\alpha_{RS}=1$ do not increase the adjusted $R^2$-value metric in Section \ref{compare}.}  

If the \textit{common parameters} are known, early measurements of $\text{PSA}(t)$ during \textit{on-} and \textit{off-drug} periods allow us to estimate the \textit{patient-specific parameters}. These estimates can be updated and refined as additional data becomes available. We can then use Eq. \eqref{prostate_cancer_small_model} to predict future $\text{PSA}(t)$ observations and the growth trajectories of the two cell types, sensitive $x_S(t)$ and resistant $x_R(t)$, in response to different drug administration schedules. 

The only explicitly time dependent term in Eq. \ref{prostate_cancer_small_model} is $\Lambda(t)$, and it lies at the heart of the clinician strategy. In principle $\Lambda(t)$ can be any function of time, yet, given the realities of delivering chemotherapy to patients, we assume that drug delivery is determined not directly by time but rather by trigger levels of measured PSA.  That is, we allow only two knobs to tune: the \textit{on-level} $\text{PSA}_\uparrow$ and \textit{off-level} $\text{PSA}_\downarrow$. In adaptive therapy, the drug is administered 
when the patient PSA level surpasses the \textit{on-level} $\text{PSA}_\uparrow$ 
and continues until it decreases to the \textit{off-level} $\text{PSA}_\downarrow$. Drug administration is then resumed only when the PSA level returns to $\text{PSA}_\uparrow$.

The clinician goal is to prevent the patient from staying in a state of developing symptoms (e.g. as urinary issues, erectile dysfunction, pelvic pain, bone pain), which emerge when their prostate cancer have advanced far enough. We denote the PSA level at which symptoms begin to appear as $\text{PSA}_\text{thr}$, a value that is \textit{not known in advance}. At that stage, drug
administration cannot be delayed without risking further complications. For all patients in \cite{zhang2022evolution}, Abiraterone was first administered to all patients well before any signs of symptoms appeared. We define the two \textit{policy parameters}, $\eta_\uparrow$ and $\eta_\downarrow$, so that $\text{PSA}_\uparrow$ is a fraction $\eta_\uparrow$ of the clinician acceptable upper limit $\text{PSA}_\text{thr}$, and $\text{PSA}_\downarrow$ is a fraction $\eta_\downarrow$ of $\text{PSA}_\uparrow$:
\begin{equation}
\text{PSA}_\uparrow \equiv \eta_\uparrow \text{PSA}_\text{thr} \ , \ \text{PSA}_\downarrow \equiv \eta_\downarrow \text{PSA}_\uparrow = \eta_\uparrow \eta_\downarrow \text{PSA}_\text{thr} \ . 
\label{trigger}
\end{equation}
The values of $\eta_\uparrow$ and $\eta_\downarrow$ are therefore bounded between $0$ and $1$.

Given this formulation, patient-informed parameters and the PSA threshold value $\text{PSA}_\text{thr}$, we can now attempt to determine what the optimal treatment policy should be for the patient.

\section{Patient-Informed Parameter Estimation \label{Extraction}}

\subsection{Clinical Data and Nonlinear Mixed Effect}

We reanalyze the clinical data from \cite{zhang2022evolution}, which includes $N=32$ patients undergoing Abiraterone treatment. Note that these patients have already received treatment to block male hormone production from the pituitary gland, which accounts for approximately $\sim 90\%$ of total production. Abiraterone is used as an additional therapy, to inhibit androgen production from other sources, including the adrenal glands and the tumor itself, which contribute the remaining $\sim 10\%$ of total production. Considerations in our analysis differ from those of \cite{zhang2022evolution} in certain {\color{black}directions}:

\begin{figure*}[!htbp]
\includegraphics[width=\textwidth]{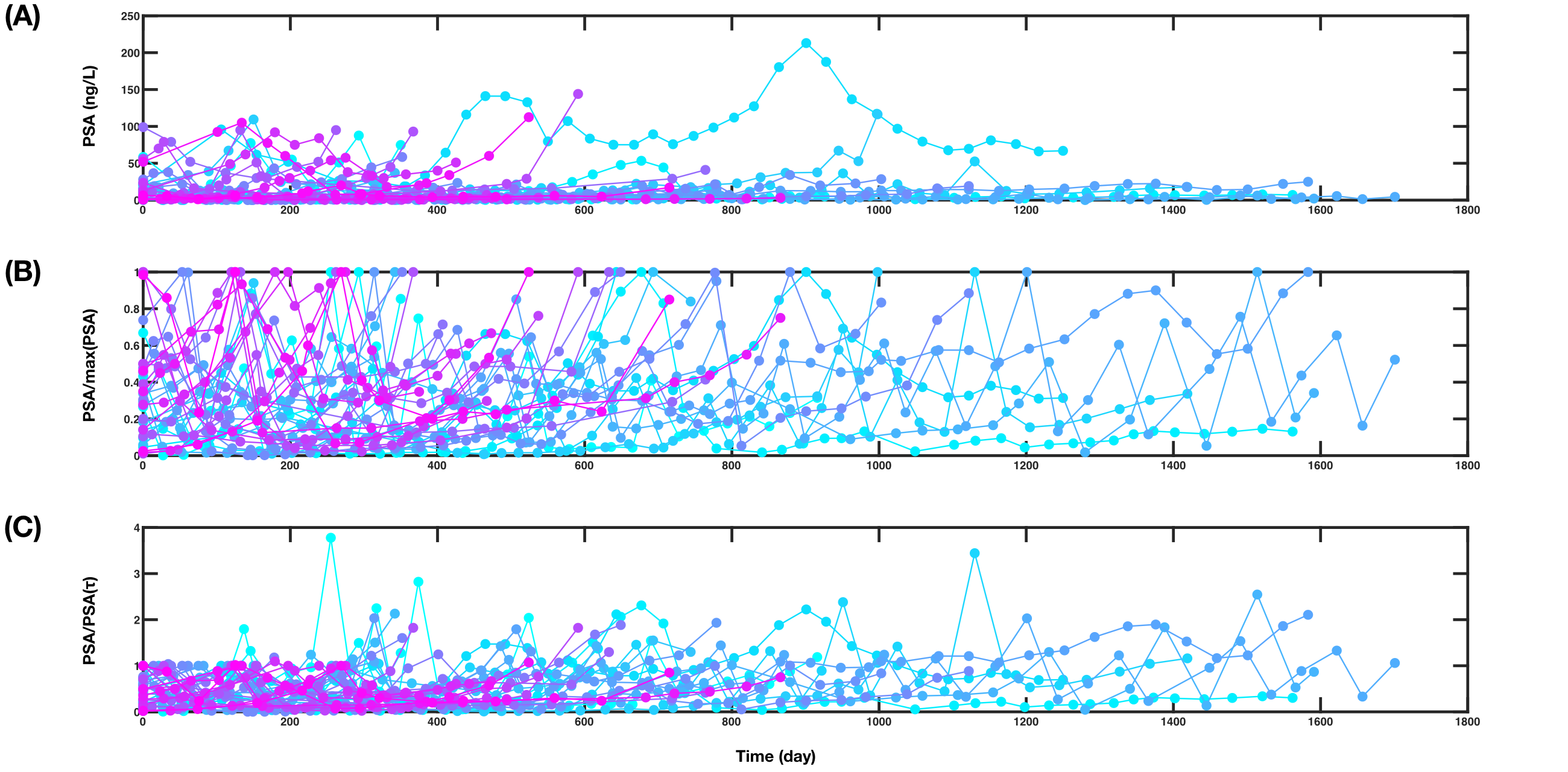}%
\caption{\justifying \textbf{Time series of PSA levels for all $N=32$ patients, for alternative normalizations}. \textbf{(A)} The PSA level in ng/L. \textbf{(B)} The PSA level as normalized in Zhang et al. \cite{zhang2022evolution}, in which the maximum value of any given PSA time series after the normalization is always at $1$. \textbf{(C)} The normalized PSA level is adjusted so that the value at the start of the first treatment in any given PSA time series is always set to $1$ after normalization.}
\label{PSA_show_off}
\end{figure*}

\begin{itemize} 
    \item In Fig. \ref{PSA_show_off}A we show the raw time series data for the PSA level (in ng/L) of all patients. It is very important to note that these data are widely different in scale (the maximum value in each of these PSA time series ranges from $\sim 2$ng/L to $\sim 200$ng/L), so before fitting the model to all patients, we need to normalize them to ensure comparability between patients. If not, the fit will then be overly controlled by the highly volatile patients. Here, we normalize each patient's PSA measurement data using the PSA value at the time Abiraterone was first administered (see Fig. \ref{PSA_show_off}C), instead of the maximum PSA value as in \cite{zhang2022evolution} (see Fig. \ref{PSA_show_off}B). This choice for normalization not only still ensures comparability between PSA time series across all patients, but also allows for data updates without necessitating changes to previous entries.
    \item We decide to relax the assumption in \cite{zhang2022evolution} that the cancer growth rates $r_S$ and $r_R$ are identical for all patients. This is consistent with the values estimated for patients receiving continuous drug treatment, ranging from $2\times10^{-3}$ to $2\times10^{-2}$/day \cite{zhang2022evolution}. Instead, we permit $r_R$ to vary among patients while treating the ratio of these rates, $r_S/r_R$, as universal. We do not pre-set $r_S/r_R$.
    \item We let the drug effectiveness $\gamma$ to be a free parameter instead of fixing it. This allows for variability in how different patients respond to the drug, and provides a more accurate assessment of how close the initial total cancer cell population is to the carrying capacity, i.e. how close $x(0)$ is to $1$.
    \item The model is nonlinear in both common as well as person specific (mixed) factors. For this reason, we utilize an established statistical approach, the \textit{empirical Bayes approach of Lindstrom and Bates} \cite{lindstrom1990nonlinear}, to fit the model to the \textit{normalized} PSA data within the framework of \textit{nonlinear mixed effects} (NLME). The estimated parameters for fitting the model to the data are selected so as to \textit{maximize the observation likelihood}, meaning they provide the best possible explanation for the patterns seen in the data. Not only do we have estimates for the two \textit{common parameters} $\{ \alpha_{RS}, r_S/r_R \}$, we also obtain probability distributions for four \textit{patient-specific parameters} 
        \begin{equation}
        \{  r_R, \gamma, x_S(0), x_R(0) \} \ .
    \label{4_ps_params}
    \end{equation}
    Estimation of the \textit{patient-specific parameters} benefits from regarding them as samples from a distribution, since it statistically addresses the regression to the mean for extreme values.
    \item We do not assign weights differently (and rather arbitrarily) to different events (e.g. the first data points after a treatment change, from off-drug to on-drug and \textit{vice versa}, are weighted five times more than the others in \cite{zhang2022evolution}). Here, all data points are treated the same.
\end{itemize}
Note that there are five \textit{patient-specific parameters} as previously mentioned in Eq. \eqref{5_ps_params}, but Eq. \eqref{4_ps_params} addresses only four: the fifth, $\tilde{\lambda}$, is determined by the other four.

To be more precise, let us denote the two \textit{common parameters} as $\Theta \equiv \{\alpha_{RS}, r_S/r_R \}$ and the four \textit{patient-specific parameters} (besides $\tilde{\lambda}$) as $\theta \equiv \{\gamma, r_R, x_S(0), x_R(0) \}$, and use the subscript $\mu$ to specify the individual patient, i.e. $\theta_\mu$. For each patient $\mu$ in \cite{zhang2022evolution}, we know their given Abiraterone treatment schedule $\Lambda_\mu (t)$, and the measurement times $t_{m_\mu}$ (where $m_\mu$ labels the time point) at which they have their PSA level recorded, as well as the measured values $\text{PSA}_\mu (t_{m_\mu})$. If $m_\mu^{\text{(treat)}}$ is the time point when Abiraterone was first administered, then the \textit{normalized} PSA level can be calculated with: 
\begin{equation}
    \overline{\text{PSA}}_\mu (t_{m_\mu}) =  \text{PSA}_\mu (t_{m_\mu})/\text{PSA}_\mu (t_{m^{\text{(treat)}}_\mu}) \ .
\label{normalized_PSA}
\end{equation}
For a given $\left\{\Theta,\theta_\mu;\Lambda_\mu(t)\right\}$, we can use Eq. \eqref{prostate_cancer_small_model} to obtain theoretically what trajectory the total cancer cell population should be,\footnote{We employ the fourth-order Runge-Kutta method for numerical integration to derive the trajectory in MatLab R2023a \cite{MATLAB}.} which we denote as $x^{\left\{\Theta,\theta_\mu;\Lambda_\mu(t)\right\}} (t)$. The theoretical trajectory for the \textit{normalized} PSA level, therefore, follows from Eq. \eqref{prostate_cancer_small_model} if $\tilde{\lambda}_\mu$ is known:
\begin{equation}
    \overline{\text{PSA}}^{\left\{\Theta,\theta_\mu;\Lambda_\mu(t)\right\}} (t) =  \tilde{\lambda}_\mu x^{\left\{\Theta,\theta_\mu;\Lambda_\mu(t)\right\}} (t) /\text{PSA}_\mu (t_{m^{\text{(treat)}}_\mu}) \ .
\end{equation}
We then estimate $\tilde{\lambda}_\mu$ by minimizing the sum squared-errors (SSE) between the observational $\overline{\text{PSA}}_\mu (t_{m_\mu})$ and the theoretical $\overline{\text{PSA}}^{\left\{\Theta,\theta_\mu;\Lambda_\mu(t)\right\}} (t_{m_\mu})$:
\begin{equation}
\begin{split}
    \tilde{\lambda}_\mu &= \text{arg} \min \left\{ \sum^{\text{given $\mu$}}_{m_\mu}\left[ \overline{\text{PSA}}_\mu (t_{m_\mu}) - \overline{\text{PSA}}^{\left\{\Theta,\theta_\mu;\Lambda_\mu(t)\right\}} (t_{m_\mu}) \right]^2 \right\}
    \\
     &= \frac{\displaystyle \sum^{\text{given $\mu$}}_{m_\mu} \overline{\text{PSA}}_\mu (t_{m_\mu}) x^{\left\{\Theta,\theta_\mu;\Lambda_\mu(t)\right\}} (t_{m_\mu})}{\displaystyle \sum^{\text{given $\mu$}}_{m_\mu} \left[ x^{\left\{\Theta,\theta_\mu;\Lambda_\mu(t)\right\}} (t_{m_\mu})\right]^2} = \frac{\displaystyle \sum^{\text{given $\mu$}}_{m_\mu} \text{PSA}_\mu (t_{m_\mu}) x^{\left\{\Theta,\theta_\mu;\Lambda_\mu(t)\right\}} (t_{m_\mu})}{\displaystyle \sum^{\text{given $\mu$}}_{m_\mu} \left[ x^{\left\{\Theta,\theta_\mu;\Lambda_\mu(t)\right\}} (t_{m_\mu})\right]^2} \ .
\end{split}
\end{equation}
In other words, $\tilde{\lambda}_\mu$ is determined by $\left\{\Theta,\theta_\mu;\Lambda_\mu(t)\right\}$ and the time series $\text{PSA}_\mu (t_{m_\mu})$ for each individual patient $\mu$.

\subsection{Statistical Assumptions}

To use the Lindstrom-Bates empirical Bayes (LBEB) approach for NLME  \cite{lindstrom1990nonlinear}, we need to make some statistical assumptions:
\begin{itemize}
    \item 
The discrepancies in the data fit are sampled from a combined-variance Gaussian noise, which includes both a constant component and a component that is proportional to the ``true'' value. This means that measurements for a larger ``true'' value are allowed to exhibit greater variability, i.e.:
\begin{equation}
   \overline{\text{PSA}}_\mu (t_{m_\mu}) = \overline{\text{PSA}}^{\left\{\Theta,\theta_\mu;\Lambda_\mu(t)\right\}} (t_{m_\mu}) + \eta_\mu (t_{m_\mu}) \ ,
\end{equation}
where the errors $\eta_\mu (t_{m_\mu})$ are sampled from a Gaussian distribution: 
\begin{equation}
    \left[ A+B \ \overline{\text{PSA}}^{\left\{\Theta,\theta_\mu;\Lambda_\mu(t)\right\}} (t_{m_\mu}) \right] \mathcal{N}(0,1) \ .
\end{equation}
Here $A$, $B$ are two unknown positive constants, and $\mathcal{N}(0,1)$ represents the Gaussian distribution centered at $0$ with a variance of $1$.
    \item The four \textit{patient-specific parameters} $\{r_R,(1-\gamma),x_S(0),x_R(0)\}$ are assumed to be sampled from four log-normal distributions \cite{crow1987lognormal}. Note that we consider $(1-\gamma)$ to be sampled from a log-normal distribution, not $\gamma$, because observations (PSA levels before and after Abiraterone administration) suggest that $\gamma$ is typically much closer to $1$ than $0$.
\end{itemize}
We then apply the LBEB approach\footnote{We use the standard package \textit{nlmefit} in MatLab R2023a \cite{MATLAB} to algorithmically 
execute LBEB approach for NLME.} to estimate the \textit{common parameters} $\{r_S/r_R, \alpha_{RS}\}$, the log-normal distributions of $\{r_R,(1-\gamma), x_S(0), x_R(0)\}$, and the values $\{r_R,\gamma, x_S(0), x_R(0)\}$ of each patient, \textit{and also} the two constants $A$, $B$ that characterized our noise function, so that the observations across patients can be the most likely ones (maximum likelihood).

\subsection{Best-Fit Results \label{Best_Fit_Est}}

\begin{figure*}[!htbp]
\includegraphics[width=\textwidth]{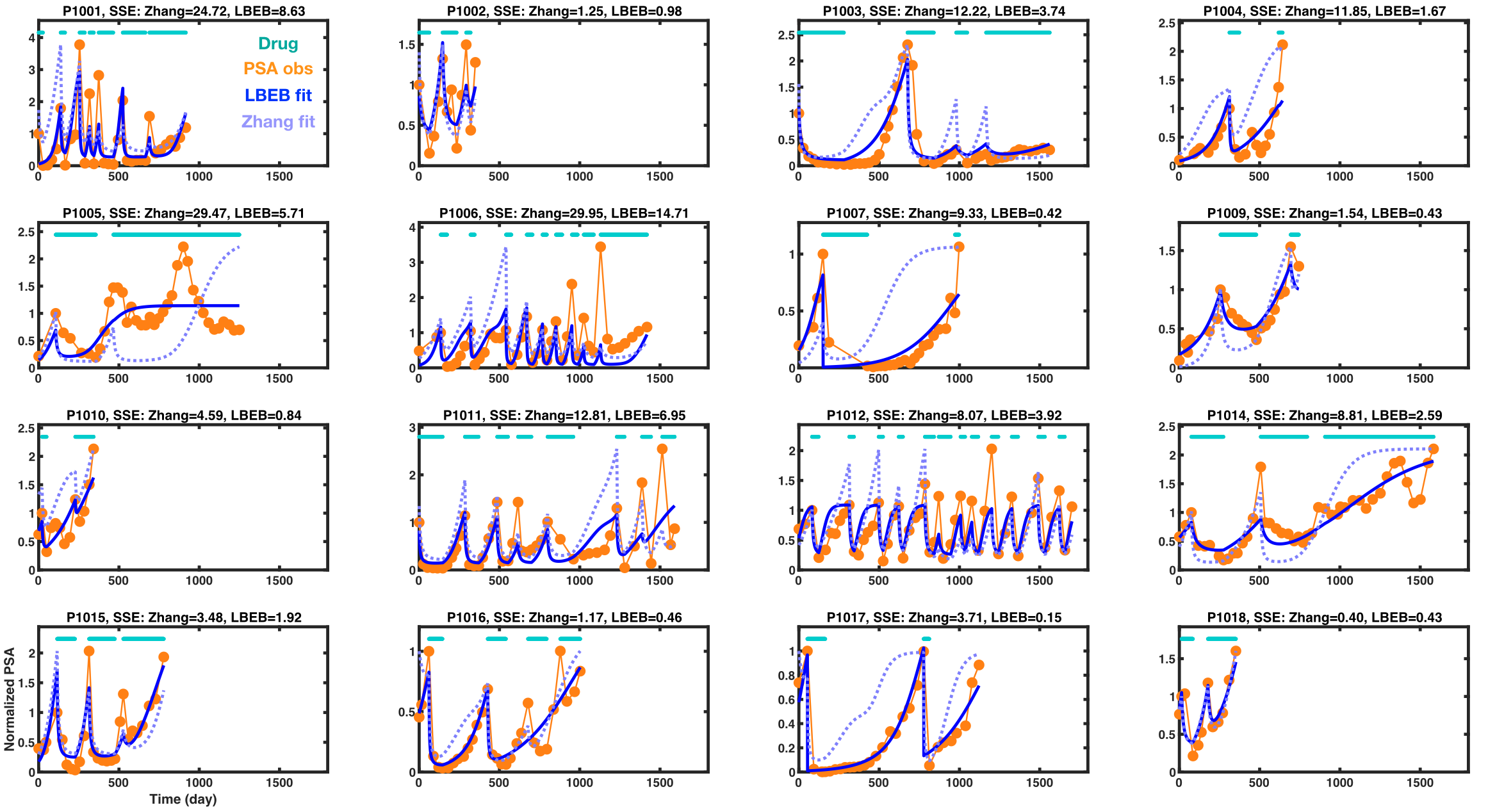}%
\caption{\justifying \textbf{Time series of PSA levels for the first 16 patients.}. Each plot title shows the patient's name, SSE for Zhang et al. \cite{zhang2022evolution} fit, and the LBEB fit. The cyan line marks drug use periods, orange dots represent normalized PSA Eq. \eqref{normalized_PSA}, the light-blue dashed line shows Zhang et al. \cite{zhang2022evolution} best fit, and the blue line shows the LBEB best fit.}
\label{First_16}
\end{figure*}

\begin{figure*}[!htbp]
\includegraphics[width=\textwidth]{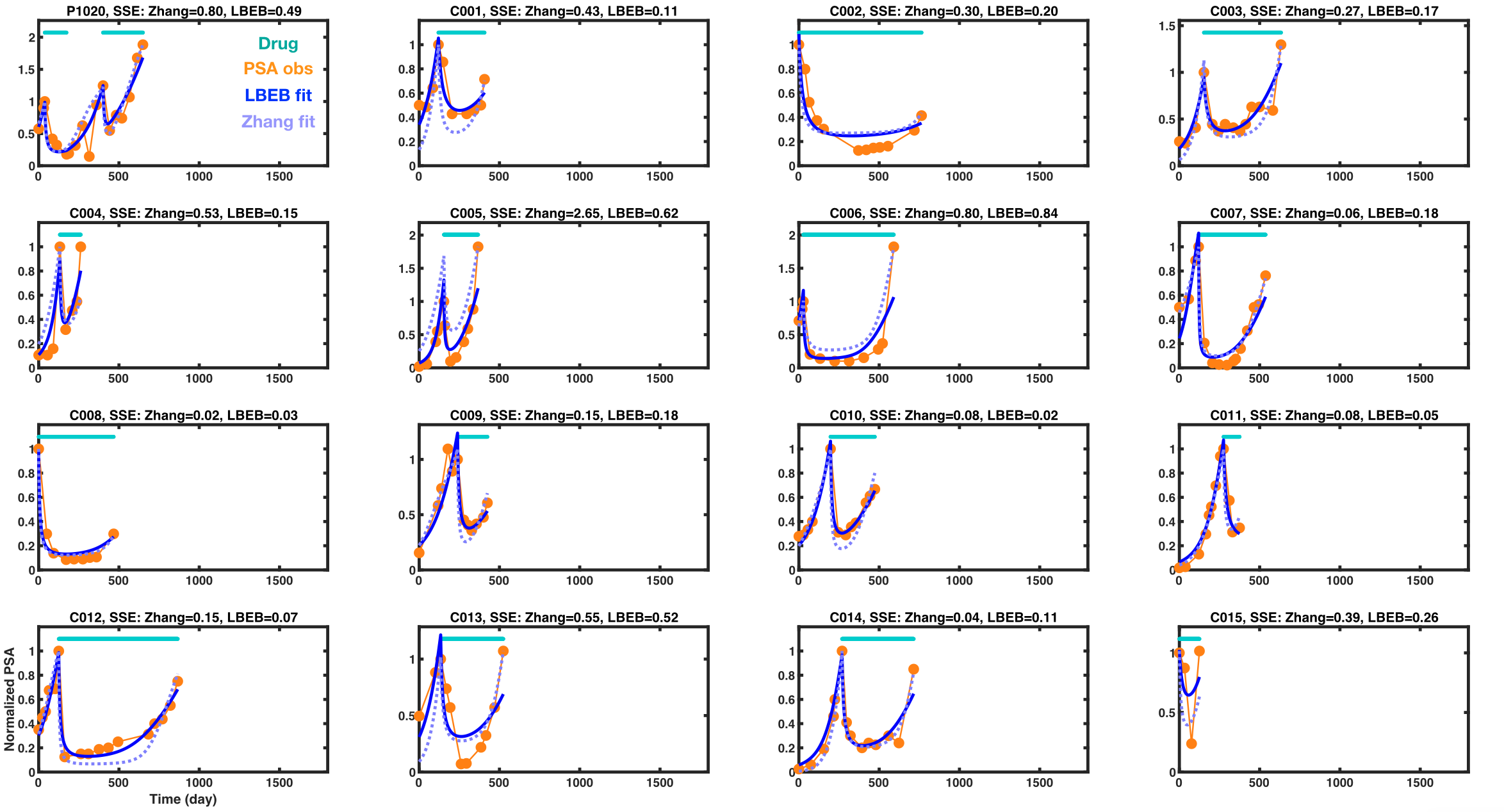}%
\caption{\justifying \textbf{Time series of PSA levels for the next 16 patients.} Each plot title shows the patient's name, SSE for Zhang et al. \cite{zhang2022evolution} fit, and the LBEB fit. The cyan line marks drug use periods, orange dots represent normalized PSA Eq. \eqref{normalized_PSA}, the light-blue dashed line shows Zhang et al. \cite{zhang2022evolution} best fit, and the blue line shows the LBEB best fit.}
\label{Second_16}
\end{figure*}

We start by acknowledging that the clinical data is extremely noisy, which should be expected. The fit found in Zhang et al. \cite{zhang2022evolution}, while not quantitative, does capture nearly all the qualitative behavior of every {\color{black}patient} regarding PSA response \textit{on-} and \textit{off-}drug. However, the total SSE for all patients of this fit is very large, which motivates us to reanalyze their data. While we still consider two \textit{common parameters}, we now use four \textit{patient-specific parameters} instead of only two, resulting in a significantly better fit, as shown in Fig. \ref{First_16} and Fig. \ref{Second_16}. The total SSE has been reduced by more than a factor of three, from $170.68$ for Zhang et al. \cite{zhang2022evolution} fit to $57.52$ for our LBEB approach. The SSE for individual patients is also typically smaller using our approach (26 out of 32).

\begin{figure*}[!htbp]
\includegraphics[width=\textwidth]{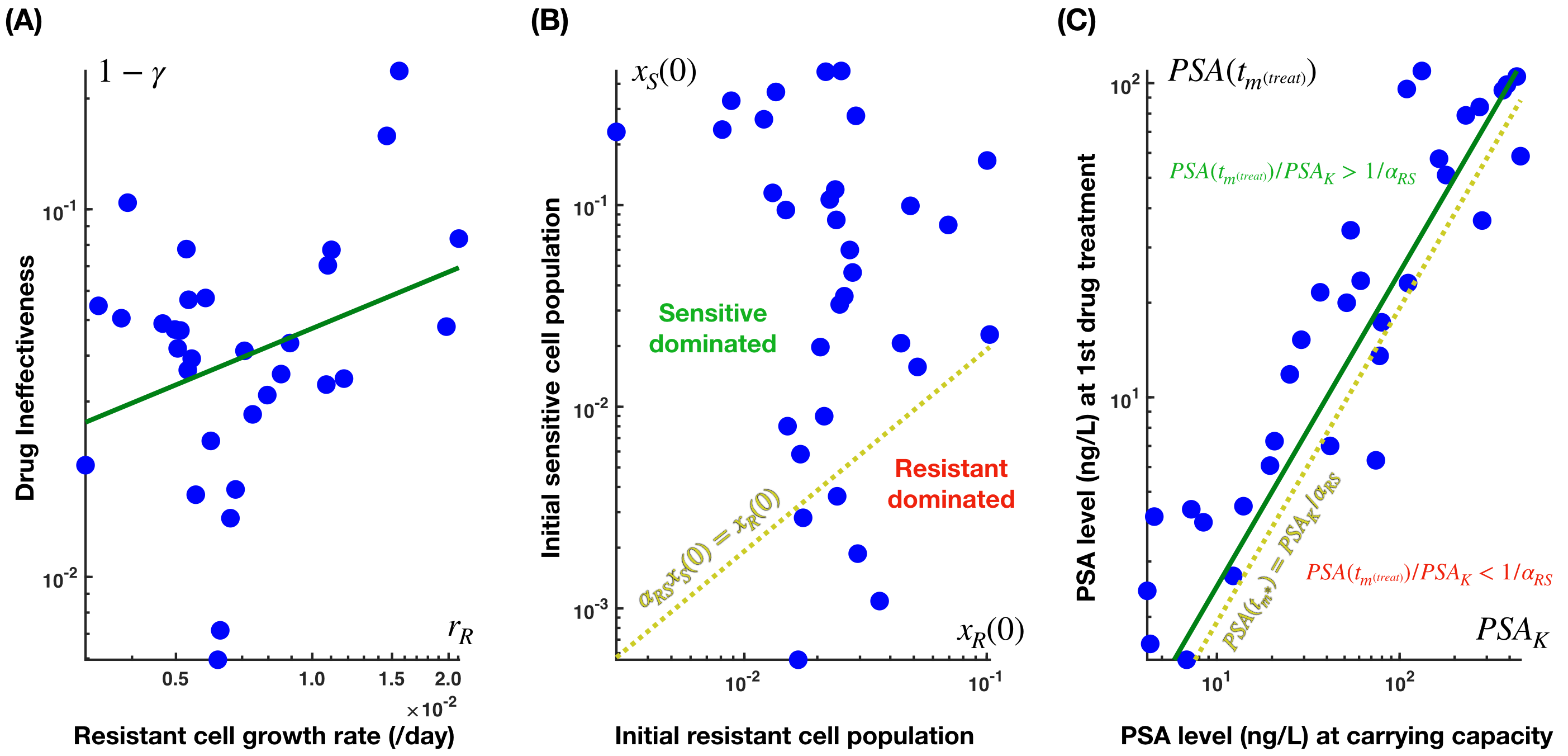}%
\caption{\justifying \textbf{The relations between patient-specific parameters and the PSA level at the first drug treatment.} Here we show the identified  resistant cell growth rate $r_R$, drug effectiveness $\gamma$, and initial population values $\{ x_S(0), x_R(0)\}$, and the PSA level when the total cancel population reaches its carrying capacity $\text{PSA}_K = \tilde{\lambda}$ for all patients.  \textbf{(A)} The resistant cell growth rate $r_R$ and the drug \textit{ineffectiveness} $1-\gamma$. The green continuous line shows a slight positive trend between these parameters in log-log plot ($R^2=0.10$ \cite{montgomery2021introduction}), indicating a power-law relation $1-\gamma \propto r_R^{0.5\pm 0.5}$. \textbf{(B)} The initial values $\{ x_S(0), x_R(0)\}$. The yellow dash line divides this two-dimensional parameter space into two: sensitive cell dominated (above) and resistant cell dominated (below), based on their influences on the growth of resistant cells at $t=0$. \textbf{(C)} The PSA levels at first drug treatment $\text{PSA}(t_{m^{\text{treat}}})$ and at carrying capacity $\text{PSA}_K$. The green continuous line shows a linear regression (that passes through the origin) between these parameters ($R^2=0.59$ \cite{montgomery2021introduction}), indicating a linear relation $\text{PSA}(t_{m^{\text{treat}}}) = \left( 0.25 \pm 0.05 \right)\text{PSA}_K $. The yellow dash line divides this two-dimensional space into two: $\text{PSA}(t_{m^{\text{treat}}})/\text{PSA}_K > 1/\alpha_{RS}$ (above) and $\text{PSA}(t_{m^{\text{treat}}})/\text{PSA}_K < 1/\alpha_{RS}$ (below).}
\label{Any_Pattern}
\end{figure*}

We find the best \textit{common parameter} values to be $r_S/r_R=2.97 \pm 0.02$ and $\alpha_{RS} = 5.2 \pm 0.9$. This provides strong evidence that $r_S > r_R$. The large value for the competition coefficient $\alpha_{RS}$ between resistant and sensitive cells agrees with the findings of \cite{zhang2022evolution}, indicating the robustness of this estimation even with different approaches. The \textit{patient-specific parameters} are sampled from $$r_R \sim (7.0 \times 10^{-3}) \exp\left[ 0.7 \ \mathcal{N}(0,1)\right] \ , \ \gamma \sim  1 - (4.0 \times 10^{-2}) \exp\left[ 1.5 \ \mathcal{N}(0,1)\right] \ , $$
$$ x_S(0) \sim (4.1 \times 10^{-2}) \exp\left[ 2.4 \ \mathcal{N}(0,1)\right] \ , \ x_R(0) \sim (2.3 \times 10^{-2}) \exp\left[ 1.8 \ \mathcal{N}(0,1)\right] \ . $$ 
We report the individual values for the {patient-specific parameters} in Appendix \ref{bestfit_params}. Fig. \ref{Any_Pattern}A shows a slight positive trend (presented in log-log plot) between the resistant cell growth rate $r_R$ and the \textit{drug ineffectiveness} $1-\gamma$. 
This indicates that in patients where cancer cells can multiply quickly and thrive, Abiraterone is usually less effective in suppressing the disease. Fig. \ref{Any_Pattern}B reveals that, initially before treatment, sensitive cells contribute the most to prohibiting the growth of resistant cells (compared to resistant cells themselves). We show the monotonic relationship between the PSA level at the time of first drug treatment, $\text{PSA}(t_{m^{\text{treat}}})$, and the PSA level at carrying capacity, $\text{PSA}_K$, in Fig. \ref{Any_Pattern}C. For $27$ out of the total $N=32$ patients, $\text{PSA}(t_{m^{\text{treat}}})/\text{PSA}_K > 1/\alpha_{RS}$, which carries significant clinical implications that we will discuss in Section \ref{can_be_chronic}.

\subsection{Comparison between Models \label{compare}}

It should, of course, be expected that with more free parameters we should achieve {\color{black}a better fit}. Therefore, we need to assess whether the improved fit justifies the additional parameters and helps avoid overfitting, ensuring that the model will perform well on new data.

For linear regression, the standard metric for ``goodness-of-fit'' is given by the $R^2$-value \cite{montgomery2021introduction}. A direct generalization of this dimensionless metric for nonlinear models \cite{efron1978regression,anderson1994model} is commonly referred to as the Efron's pseudo-$R^2$-value \cite{schabenberger2001contemporary}, is given by $\tilde{R}^2 = 1-\tilde{A}$. The \textit{coefficient of alienation} $\tilde{A}$, defined as \cite{anderson1994model}:
\begin{equation} 
\tilde{A}=\frac{\text{SSE}}{\text{SS}_{\text{data}}} \geq 0 \ ,
\label{Efron_A}
\end{equation}
where $\text{SS}_{\text{data}}$ is the data variance (i.e. SSE with respect to the mean value). This measures the proportion of residual variation in the nonlinear model compared to a null model with no explanatory power, indicating how much the model deviates from an ideal fit. An ideal fit corresponds to $\tilde{R}^2=1$. While there are many other different proposals for goodness-of-fit depending on the specific characteristics of the nonlinear model being used \cite{mcfadden1972conditional, smith2013comparison,spiess2010evaluation}, the Efron's pseudo-$\tilde{R^2}$-value remains a useful metric due to its simplicity and ease of interpretation \cite{anderson1994model}.

Since the number of parameters in the fit models that we use are not the same, i.e. $n_{params}=(2\times N)+2=66$ in Zhang et al. \cite{zhang2022evolution} and $n_{params}=(4\times N)+2=130$ in our approach, we need to calculate the \textit{adjusted} Efron's psuedo-$R^2$-value \cite{archontoulis2015nonlinear} instead:
\begin{equation}
    \tilde{R}^2_{adj} = 1 - \frac{\text{SSE}/(n_{\text{data}}-n_{\text{params}})}{\text{SS}_{\text{data}}/(n_{\text{data}}-1)} \ .
\end{equation}
We have $n_{\text{data}}=678$ and $\text{SS}_{\text{data}}=176.16$. The closer $\tilde{R}^2_{adj}$ is to $1$, the more we can trust a fitted model. We obtain $\tilde{R}^2_{adj}\approx-0.08$ for Zhang et al. \cite{zhang2022evolution} fit and $\tilde{R}^2_{adj}\approx0.60$ for our LBEB fit, which corroborates our approach. A similar conclusion is reached using the Akaike information criterion (AIC) and Bayesian information criterion (BIC) analyses \cite{akaike1974new,schwarz1978estimating}, i.e.
\begin{itemize}
    \item $\mathrm{AIC}_{\text{Zhang}} \approx -798 > \mathrm{AIC}_{\text{LBEB}} \approx -1418$, where $\mathrm{AIC} = n_{\text{data}} \ln \left( \mathrm{SSE}/n_{\text{data}} \right) + 2n_{\text{params}}$.
    \item $\mathrm{BIC}_{\text{Zhang}} \approx -500 > \mathrm{BIC}_{\text{LBEB}} \approx -831$, where $\mathrm{BIC} = n_{\text{data}} \ln \left( \mathrm{SSE}/n_{\text{data}} \right) + n_{\text{params}} \ln (n_{\text{data}})$.
\end{itemize}

\section{Optimization of Treatment with Abiraterone \label{Optimization}}

Within our adaptive therapy policy framework, our goal is to determine the optimal policy parameters $(\eta_\uparrow, \eta_\downarrow)$ that maximize the \textit{total achievable time} $\tau_{\text{tot}}$ we can keep the cancer progression (with treatment) below the biomarker level $\text{PSA}_{\text{thr}}$, the level at which the patient begins to show symptoms, in order to avoid further complications. Consider the patients with two \textit{common parameters} $\Theta = \{ r_S/r_R, \alpha_{RS}\}$ and the five \textit{patient-specific parameters} $\{\text{PSA}_K,\theta \}=\{ \tilde{\lambda}, r_R, \gamma, x_S(0), x_R(0) \}$, as found according to Section \ref{Best_Fit_Est}. Note that $\text{PSA}_K=\tilde{\lambda}$ follows, as explained in Eq. \eqref{PSA_K}. For simplification, we further assume that the PSA level is measured continuously, and that there is no noise in these measurements, so that the decision to turn \textit{on} and \textit{off} drug treatment based on the policy parameters as described in Eq. \eqref{trigger} is exact. Hence, we can calculate the \textit{total achievable time} $\tau_{\text{tot}}$ below $\text{PSA}_{\text{thr}}$ using policy parameters $(\eta_\uparrow, \eta_\downarrow)$ for a patient $\mu$ with parameters $\{ {\Theta, \text{PSA}_K}_\mu, \theta_\mu \}$,\footnote{We utilize the standard integrator \textit{ode23} in MatLab R2023a \cite{MATLAB} to simulate Eq. \eqref{prostate_cancer_small_model} and implement the terminal condition, obtaining the \textit{total achievable time} $\tau_{\text{tot}}$ at when the simulated PSA level crosses the threshold value $\text{PSA}_{\text{thr}}$.} and we denote its full functional form as $\tau_{\text{tot}}^{\{ {\Theta, \text{PSA}_K}_\mu, \theta_\mu \}, \text{PSA}_{\text{thr}}}(\eta_\uparrow, \eta_\downarrow)$. We can simplify this expression to $\tau_{\text{tot}}^{\{\mu \},\text{PSA}_{\text{thr}}}(\eta_\uparrow, \eta_\downarrow)$, and further drop all superscripts when it is clear which patient and what threshold are referred to.

As we will show in this Section through computer-assisted modeling, if the PSA threshold is too low $\text{PSA}_{\text{thr}}\leq {\text{PSA}_K}_\mu/\alpha_{RS}$ (remember that $1/\alpha_{RS} \approx 0.19$), then for any policy parameters $(\eta_\uparrow, \eta_\downarrow)$, the treatment appears to eventually fail, i.e. hence we can only hope to somewhat prolong the drug response \textit{total achievable time} $\tau_{\text{tot}}$. However, if this threshold is just high enough $\text{PSA}_{\text{thr}}>{\text{PSA}_K}_\mu/\alpha_{RS}$, then sometimes, \textit{but not always}, it is possible to find a treatment policy that arrests the cancer progression indefinitely $\tau_{\text{tot}} \rightarrow \infty$. We provide some analytical explanation for these possibilities in Section \ref{analytic}.

\subsection{$\text{PSA}_{\text{thr}}/{\text{PSA}_K}_\mu \leq 1/\alpha_{RS}$}

Here, we search for the (global) maximum of the \textit{total achievable time} (with treatment) below the PSA threshold, $\tau_{\text{tot}}^{\{\mu \},\text{PSA}_{\text{thr}}}(\eta_\uparrow, \eta_\downarrow)$ over this multi-dimensional space $(\eta_\uparrow,\eta_\downarrow)$. We employ a Bayesian optimization approach \cite{frazier2018tutorial} to efficiently navigate through the policy parameter space by data-driven extrapolation ({\color{black}a landscape-estimation method known as} kriging \cite{oliver1990kriging}) for the two-dimensional landscape $\tau_{\text{tot}}^{\{\mu \},\text{PSA}_{\text{thr}}}(\eta_\uparrow, \eta_\downarrow)$ for patient $\mu$ with a given $\text{PSA}_\text{thr}$. This method, when applied to an objective function to be {\color{black} extremized}, begins by evaluating that function at a number of random positions so as to (approximately) gauge its global features (using a surrogate model e.g. a Gaussian process \cite{schulz2018tutorial}). It then uses Bayesian inference to predict the optimal next function evaluation location, balancing exploitation and exploration, and evaluate the function there. This iterative step continuously updates its estimation of the function, keeps finding and checking new locations and corresponding values, until the difference between two consecutive locations becomes sufficiently small (for a practical resolution of the parameter space).

\subsubsection{Bayesian Optimization \label{BO}}

We briefly explain how we perform the computation for our optimal policy parameters
\begin{equation}
    (\eta_\uparrow^*,\eta_\downarrow^*) = \text{arg} \max \left[ \tau_{\text{tot}}^{\{\mu \},\text{PSA}_{\text{thr}}}(\eta_\uparrow, \eta_\downarrow)\right] \ ,
\end{equation}
using Bayesian optimization \cite{frazier2018tutorial} with Gaussian process regression (GPR) \cite{schulz2018tutorial}. Consider a search within the region $(\eta_\uparrow,\eta_\downarrow) \in [0,0.90] \times [0,0.90]$, where the resolution in each direction is $\delta \eta = 0.01$. Note that the upper bounds for $\eta_\uparrow$ and $\eta_\downarrow$ are both set at $0.90$ instead of $1.00$, as smaller differences (less that $10\%$) are clinically difficult to monitor \cite{hansen-read}. We start by choosing $n_{\text{eval}}=10$ random points 
$$\left\{ (\eta_\uparrow^{(1)},\eta_\downarrow^{(1)}), (\eta_\uparrow^{(2)},\eta_\downarrow^{(2)}), ..., (\eta_\uparrow^{(n_\text{eval})},\eta_\downarrow^{(n_\text{eval})}) \right\}$$ 
within the search range of $(\eta_\uparrow,\eta_\downarrow)$, and each point is assessed to determine the value of the \textit{total achievable time} $\tau_{\text{tot}}^{\{\mu \},\text{PSA}_{\text{thr}}}(\eta_\uparrow, \eta_\downarrow)$. Using these initial evaluations, we then employ GPR to create a probabilistic extrapolation of $\tau_{\text{tot}}^{\{\mu \},\text{PSA}_{\text{thr}}}(\eta_\uparrow, \eta_\downarrow)$ from the set of known values 
$$\left\{\tau_{\text{tot}}^{\{\mu \},\text{PSA}_{\text{thr}}}(\eta_\uparrow^{(1)},\eta_\downarrow^{(1)}), \tau_{\text{tot}}^{\{\mu \},\text{PSA}_{\text{thr}}}(\eta_\uparrow^{(2)},\eta_\downarrow^{(2)}), ..., \tau_{\text{tot}}^{\{\mu \},\text{PSA}_{\text{thr}}}(\eta_\uparrow^{(n_\text{eval})},\eta_\downarrow^{(n_\text{eval})})\right\} \ . $$
This set provides not only an estimation 
$$\hat{\tau}_{\text{tot}}^{\{\mu \},\text{PSA}_{\text{thr}};\left\{ (\eta_\uparrow^{(1)},\eta_\downarrow^{(1)}),(\eta_\uparrow^{(2)},\eta_\downarrow^{(2)}), ..., (\eta_\uparrow^{(n_\text{eval})},\eta_\downarrow^{(n_\text{eval})}) \right\}}(\eta_\uparrow, \eta_\downarrow)$$
for the \textit{total achievable time} at any untested point, but also quantifies the uncertainty of this estimation, i.e. the standard deviation
$$\hat{\sigma}_{\hat{\tau}_{\text{tot}}}^{\{\mu \},\text{PSA}_{\text{thr}};\left\{ (\eta_\uparrow^{(1)},\eta_\downarrow^{(1)}), (\eta_\uparrow^{(2)},\eta_\downarrow^{(2)}), ..., (\eta_\uparrow^{(n_\text{eval})},\eta_\downarrow^{(n_\text{eval})}) \right\}}(\eta_\uparrow, \eta_\downarrow) \ .$$
Bayesian optimization {\color{black}utilizes} this GPR extrapolation\footnote{We utilize the function \textit{fitrgp} in MatLab R2023a \cite{MATLAB} for this task, using the \textit{ardmatern32} kernel function and the ``exact fitting'' option.} to identify the next candidate points in the search space, $(\eta_\uparrow^{(n_\text{eval}+1)},\eta_\downarrow^{(n_\text{eval}+1)})$, where the \textit{total achievable time} is likely to be maximized with some confidence. This is done by balancing exploration (searching in regions with high uncertainty) and exploitation (focusing on regions predicted to have high values):
\begin{equation}
    \begin{split}(\eta_\uparrow^{(n_\text{eval}+1)},\eta_\downarrow^{(n_\text{eval}+1)}) = \text{arg} \max \Bigg[ \hat{\tau}_{\text{tot}}^{\{\mu \},\text{PSA}_{\text{thr}};\left\{ (\eta_\uparrow^{(1)},\eta_\downarrow^{(1)}), (\eta_\uparrow^{(2)},\eta_\downarrow^{(2)}), ..., (\eta_\uparrow^{(n_\text{eval})},\eta_\downarrow^{(n_\text{eval})}) \right\}}(\eta_\uparrow, \eta_\downarrow) &
    \\
     + \chi \hat{\sigma}_{\hat{\tau}_{\text{tot}}}^{\{\mu \},\text{PSA}_{\text{thr}};\left\{ (\eta_\uparrow^{(1)},\eta_\downarrow^{(1)}), (\eta_\uparrow^{(2)},\eta_\downarrow^{(2)}), ..., (\eta_\uparrow^{(n_\text{eval})},\eta_\downarrow^{(n_\text{eval})}) \right\}}(\eta_\uparrow, \eta_\downarrow)&  \Bigg] \ .
    \end{split}
\end{equation}
We consider $\chi=2$, corresponding to $\approx 95\%$ confidence. We then update the extrapolation and repeat the above procedure for subsequent iterations, i.e. $n\geq n_{\text{eval}}$:
\begin{equation}
\begin{split}
    &\hat{\tau}_{\text{tot}}^{\{\mu \},\text{PSA}_{\text{thr}};\left\{ (\eta_\uparrow^{(1)},\eta_\downarrow^{(1)}),(\eta_\uparrow^{(2)},\eta_\downarrow^{(2)}), ..., (\eta_\uparrow^{(n)},\eta_\downarrow^{(n)}) \right\}}(\eta_\uparrow, \eta_\downarrow)
    \\
    & \ \ \longrightarrow \ \ \hat{\tau}_{\text{tot}}^{\{\mu \},\text{PSA}_{\text{thr}};\left\{ (\eta_\uparrow^{(1)},\eta_\downarrow^{(1)}),(\eta_\uparrow^{(2)},\eta_\downarrow^{(2)}), ..., (\eta_\uparrow^{(n+1)},\eta_\downarrow^{(n+1)}) \right\}}(\eta_\uparrow, \eta_\downarrow) \ , 
\end{split}
\end{equation}
\begin{equation}
\begin{split}
    &\hat{\sigma}_{\hat{\tau}_{\text{tot}}}^{\{\mu \},\text{PSA}_{\text{thr}};\left\{ (\eta_\uparrow^{(1)},\eta_\downarrow^{(1)}),(\eta_\uparrow^{(2)},\eta_\downarrow^{(2)}), ..., (\eta_\uparrow^{(n)},\eta_\downarrow^{(n)}) \right\}}(\eta_\uparrow, \eta_\downarrow)
    \\
    & \ \ \longrightarrow \ \ \hat{\sigma}_{\hat{\tau}_{\text{tot}}}^{\{\mu \},\text{PSA}_{\text{thr}};\left\{ (\eta_\uparrow^{(1)},\eta_\downarrow^{(1)}),(\eta_\uparrow^{(2)},\eta_\downarrow^{(2)}), ..., (\eta_\uparrow^{(n+1)},\eta_\downarrow^{(n+1)}) \right\}}(\eta_\uparrow, \eta_\downarrow) \ ; 
\end{split}
\end{equation}
we then calculate: 
\begin{equation}
    \begin{split}(\eta_\uparrow^{(n+2)},\eta_\downarrow^{(n+2)}) = \text{arg} \max \Bigg[ \hat{\tau}_{\text{tot}}^{\{\mu \},\text{PSA}_{\text{thr}};\left\{ (\eta_\uparrow^{(1)},\eta_\downarrow^{(1)}), (\eta_\uparrow^{(2)},\eta_\downarrow^{(2)}), ..., (\eta_\uparrow^{(n+1)},\eta_\downarrow^{(n+1)}) \right\}}(\eta_\uparrow, \eta_\downarrow) &
    \\
     + \chi \hat{\sigma}_{\hat{\tau}_{\text{tot}}}^{\{\mu \},\text{PSA}_{\text{thr}};\left\{ (\eta_\uparrow^{(1)},\eta_\downarrow^{(1)}), (\eta_\uparrow^{(2)},\eta_\downarrow^{(2)}), ..., (\eta_\uparrow^{(n+1)},\eta_\downarrow^{(n+1)}) \right\}}(\eta_\uparrow, \eta_\downarrow)&  \Bigg] \ .
    \end{split}
\end{equation}
This iterative refinement continues until satisfactory convergence is achieved (which is chosen based on the resolution $\delta\eta=0.01$), and the final iteration gives an estimated value for $(\eta_\uparrow^*,\eta_\downarrow^*)$ as the discovered global maximum of $\tau_{\text{tot}}^{\{\mu \},\text{PSA}_{\text{thr}}}(\eta_\uparrow, \eta_\downarrow)$  along with the associated uncertainty. 

\subsubsection{Optimal Treatment for Maximum $\tau_{\text{tot}}$ \label{optim}}

\begin{figure*}[!htbp]
\includegraphics[width=\textwidth]{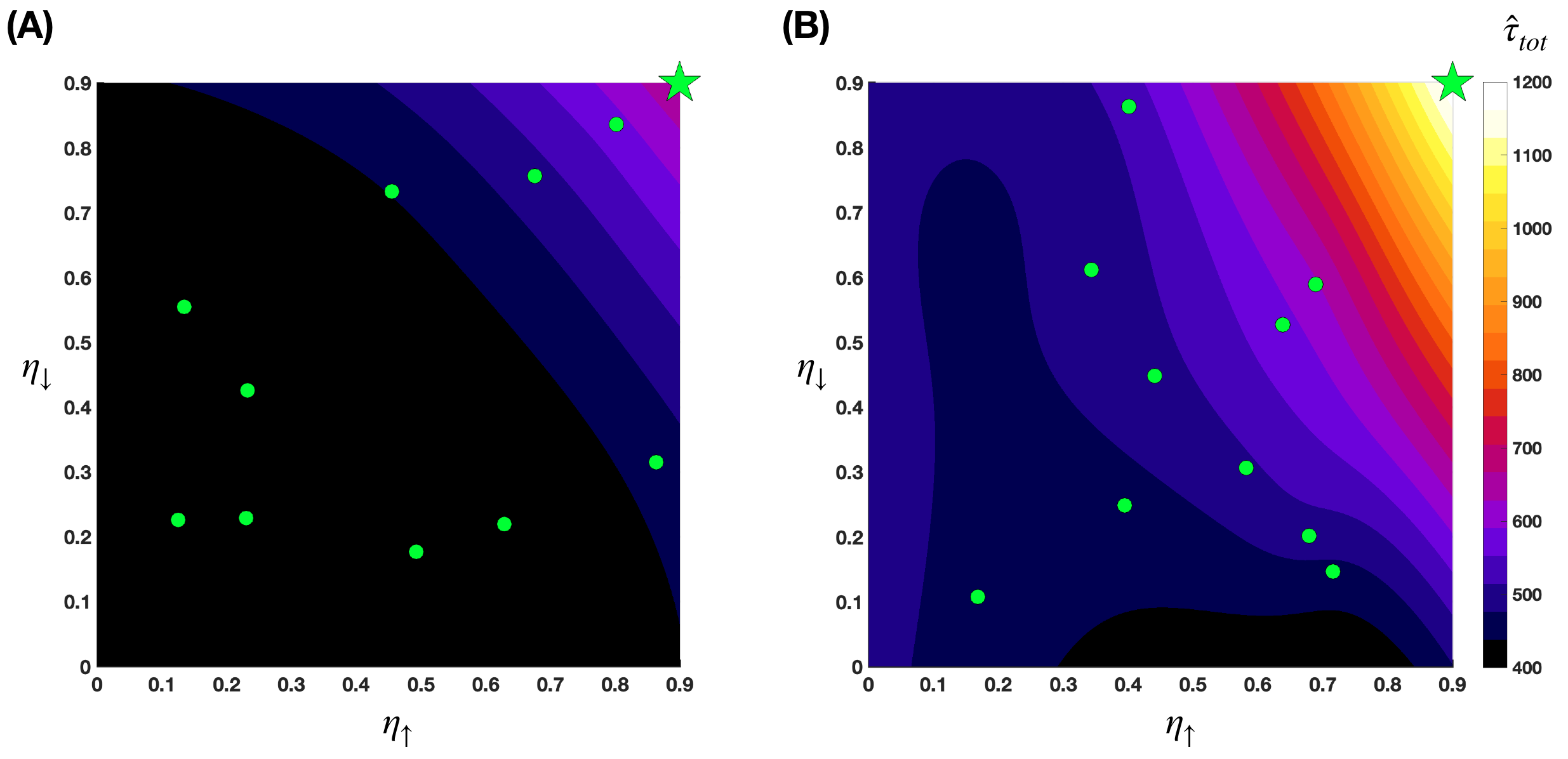}%
\caption{\justifying \textbf{Bayesian optimization using a GPR model to search for the maximum total drug response time
$\tau_{\text{tot}}$ for patient P1007 under adaptive chemotherapy.} We consider different PSA threshold values, i.e. $\text{PSA}_\text{thr}/\text{PSA}_K=0.14$ in \textbf{(A)} and $\text{PSA}_\text{thr}/\text{PSA}_K=0.19$. Here we show the Gaussian extrapolated surfaces of the \textit{total achievable time}, as given in Eq. \eqref{GPR_tau}. The circles represent the points where the \textit{total achievable time} is evaluated by numerical simulation. The
star indicates the point of convergence for the search, which corresponds to the global {\color{black}maximum}.}
\label{surface}
\end{figure*}

 The optimization described in Section \ref{BO} can be applied to any patient; however, for illustrative purposes, we will consider patient P1007 as labelled in Fig. \ref{First_16}. This patient has small values for $x_S(0)$ and $x_R(0)$, with $x_S(0)\gg x_R(0)$, allowing them to undergo several drug treatment cycles. We investigate $\tau_\text{tot}(\eta_\uparrow,\eta_\downarrow)$ for $\text{PSA}_\text{thr}/\text{PSA}_K=0.14$ and $0.19$. Fig. \ref{surface} shows the GPR extrapolated surface functions 
 \begin{equation}
 \hat{\tau}_\text{tot}^{\{\mu \},\text{PSA}_{\text{thr}};\left\{ (\eta_\uparrow^{(1)},\eta_\downarrow^{(1)}),(\eta_\uparrow^{(2)},\eta_\downarrow^{(2)}), ..., (\eta_\uparrow^*,\eta_\downarrow^*) \right\}}(\eta_\uparrow,\eta_\downarrow)
 \label{GPR_tau}
 \end{equation}
 for these different PSA thresholds, i.e. $\text{PSA}_\text{thr}/\text{PSA}_K=0.14$ in Fig. \ref{surface}A and $\text{PSA}_\text{thr}/\text{PSA}_K=0.19$ in Fig. \ref{surface}B. The optimal policy parameter values $(\eta^*_\uparrow,\eta^*_\downarrow)$ is located at the corner $(0.90,0.90)$, corresponding to the \textit{highest possible} \textit{on}-drug PSA level with the \textit{tightest control} ($\eta_\downarrow$ is maximum). We will see that this is consistent with our analytical considerations in Section \ref{analytic}.

Fig. \ref{Adaptive} shows our numerical findings for simulated treatments on this patient, using different policy parameters:

\begin{figure*}[!htbp]
\includegraphics[width=\textwidth]{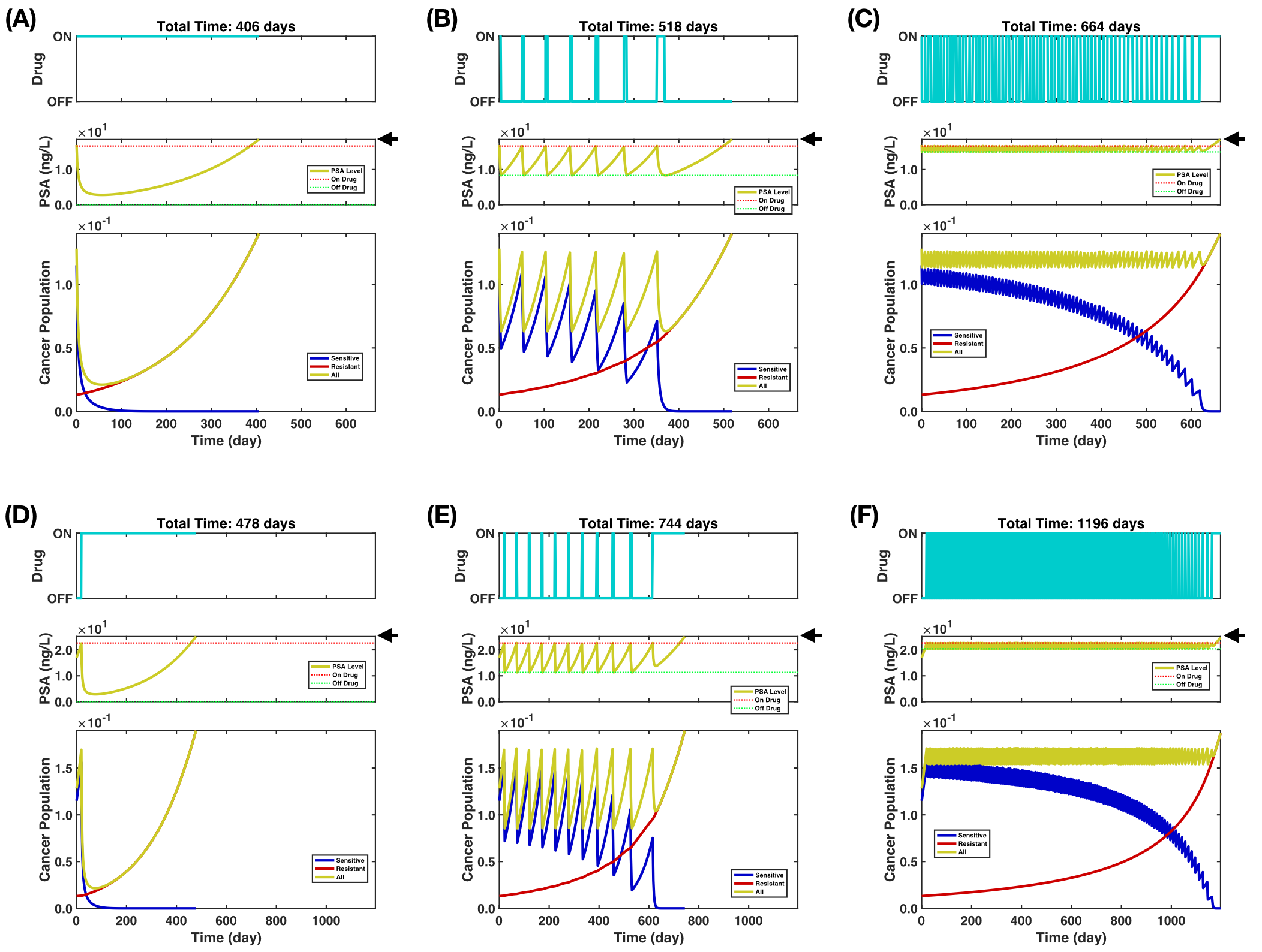}%
\caption{\justifying \textbf{Slowing down cancer progression.} We explore the dynamics of cancer progression under different treatments, for small PSA thresholds i.e. $\text{PSA}_\text{thr}/\text{PSA}_K=0.14$ in \textbf{(A,B,C)} and $0.19$ in \textbf{(D,E,F)}. We show the response of patient P1007 under a continuous therapy with $\eta_\uparrow=0.90$ in \textbf{(A,D)}, under a continuous therapy with $(\eta_\uparrow,\eta_\downarrow)=(0.90,0.50)$ in \textbf{(B,E)}, and under a continuous therapy with $(\eta_\uparrow,\eta_\downarrow)=(0.90,0.90)$ in \textbf{(C,F)}. The black arrow indicates the PSA threshold level for each plot.}
\label{Adaptive}
\end{figure*}
 
\begin{itemize}
    \item Panel (A,D): Cancer progression under continuous Abiraterone treatment often results in resistant cells rapidly overtaking the population, causing the cancer to exceed the thresholds faster. Note that continuous therapy is a special case of adaptive therapy in which the \textit{off}-drug policy parameter $\eta_{\downarrow}$ is set to $0$.
    \item Panel (B,E): Cancer progression under a (non-optimal) adaptive Abiraterone treatment, in which $\text{PSA}_\uparrow$ is as high as possible and $\text{PSA}_\downarrow$ is set to $50\%$ of $\text{PSA}_\uparrow$. This treatment has been used in studies and clinical trials \cite{zhang2022evolution}. Since the drug is not used to suppress sensitive cells constantly, these cells are allowed to recover and keep competing with the resistant ones longer, hence the \textit{total achievable time} is extended.
    \item Panel (C,F): Cancer progression under the optimal adaptive treatment, found from Bayesian optimization. This suggests that, the closer the PSA level at which the patient starts showing symptoms $\text{PSA}_\text{thr}$ is to $\text{PSA}_K/\alpha_{RS}$ (which is rather far from when the cancer population may reach carrying capacity, since $\alpha_{RS}\gg 1$), the more effective optimal adaptive therapy can be in prolonging the time gained compared to continuous treatment.

\end{itemize}
We observe in Fig. \ref{surface} that the most effective adaptive therapies maintain \textit{tight control} of PSA levels with the \textit{highest} \textit{on}-drug PSA level, i.e. $\eta_\downarrow, \eta_\uparrow \rightarrow 1$, a strategy we refer to as \textit{high level tight control} (HLTC). This treatment cycle involves administering Abiraterone in short bursts, followed by extended resting periods, at a high frequency. We report our found optimal policy parameters $(\eta_\uparrow^*,\eta_\downarrow^*)$ for all patients with $x(0)<\text{PSA}_{\text{thr}}/\text{PSA}_K$ in the Appendix \ref{report}, most of those are also $(0.90,0.90)$. We will provide an analytical justification for this finding in {\color{black}Appendix B.}

\subsection{$\text{PSA}_{\text{thr}}/{\text{PSA}_K}_\mu>1/\alpha_{RS}$ \label{can_be_chronic}}

\begin{figure*}[!htbp]
\includegraphics[width=\textwidth]{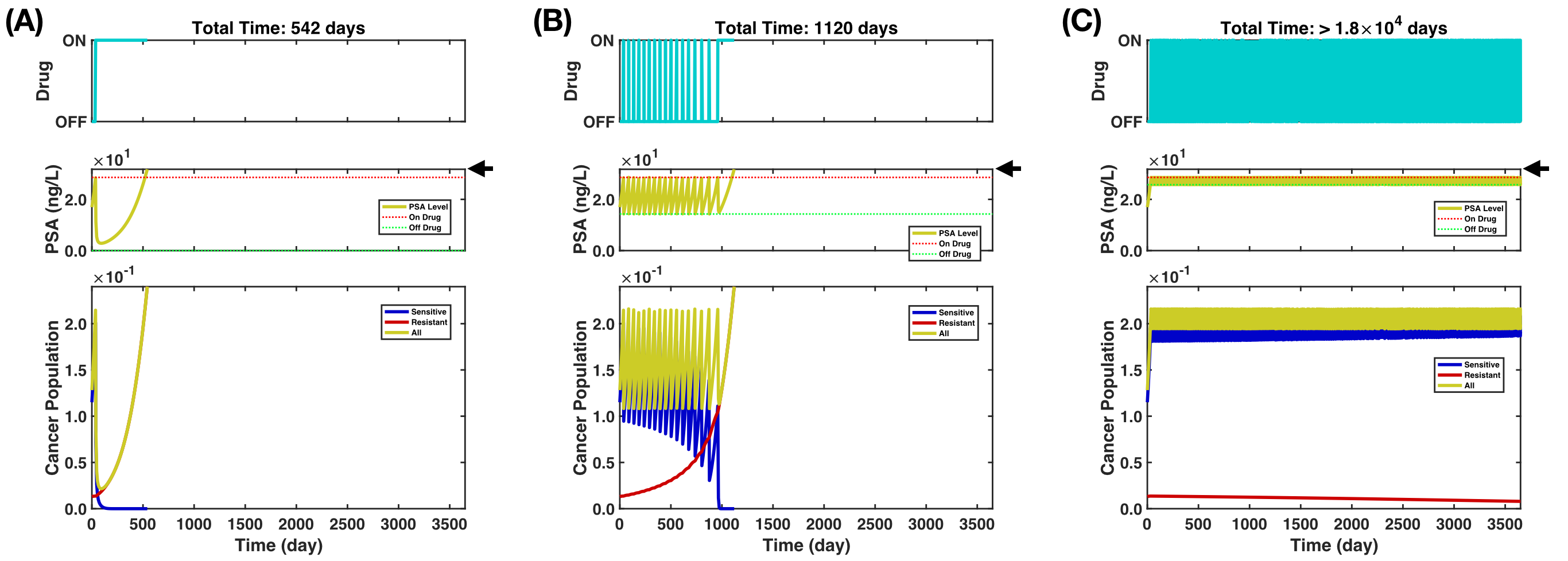}%
\caption{\justifying \textbf{Chronic cancer progression.} We explore the long-time dynamics of cancer progression under different treatments, for a large PSA threshold i.e. $\text{PSA}_\text{thr}/\text{PSA}_K=0.24$. We show the response of patient P1007 under a continuous therapy with $\eta_\uparrow=0.90$ in \textbf{(A)}, under an adaptive therapy with $(\eta_\uparrow,\eta_\downarrow)=(0.90,0.50)$ in \textbf{(B)}, and under adaptive therapy with $(\eta_\uparrow,\eta_\downarrow)=(0.90,0.90)$ in \textbf{(C)}. The black arrow indicates the PSA threshold for each plot.}
\label{Chronic}
\end{figure*}

Out of $N=32$ patients, $27$ of them have $\text{PSA}(t_{m^{\text{treat}}_\mu})/{\text{PSA}_K}_\mu>1/\alpha_{RS}$, where $\text{PSA}(t_{m^{\text{treat}}_\mu})$ is the PSA {\color{black}level} of patient $\mu$ when Abiraterone is first administrated (see Fig. \ref{Any_Pattern}C). Since $\text{PSA}(t_{m^{\text{treat}}_\mu}) < \text{PSA}_{\text{thr}}$ (all patients show no symptom at $t=t_{m^{\text{treat}}_\mu}$), this {\color{black}means} that the condition $\text{PSA}_{\text{thr}}/{\text{PSA}_K}_\mu>1/\alpha_{RS}$ is commonly satisfied.

While the \textit{total achievable time} $\tau_\text{tot}(\eta_\uparrow,\eta_\downarrow)$ is always finite for $\text{PSA}_{\text{thr}}<{\text{PSA}_K}_\mu/\alpha_{RS}$, as resistant cells will eventually dominate the cancer population after many drug administration cycles, for $\text{PSA}_{\text{thr}}>{\text{PSA}_K}_\mu/\alpha_{RS}$ it might become possible to keep the number of resistant cells under control forever. This can be realized only when the number of sensitive cells can reduce faster than the resistant cells can {\color{black}increase} during the \textit{on}-drug period (i.e. $dx_S/dt + dx_R/dt < 0$), as we will derive in Appendix \ref{analytic}. In other words, cancer can then shift from being a lethal disease to a  chronic condition. In Fig. \ref{Chronic}, we explore the consequences of $\text{PSA}_\text{thr}/\text{PSA}_K=0.24$ for the same patient as in Section \ref{optim}, on cancer progression under adaptive chemotherapy over a long time period. We find that a HLTC treatment can extend the \textit{total achievable time} indefinitely (we have checked up to $50$ years $\approx 1.8\times 10^4$ days), as illustrated in Fig. \ref{Chronic}C. However, if the PSA \text{off}-level $\text{PSA}_\downarrow$ is too low, a threshold breach can still occur, as seen in Fig. \ref{Chronic}B. Continuous treatment appears to always result in negative outcomes, as shown in Fig. \ref{Chronic}A.

What we have observed in this Section and in Section \ref{optim} is that HLTC treatment typically leads to improved outcomes, either by maximizing the total drug response time for low $\text{PSA}_{\text{thr}}$ or by transforming the disease into a chronic condition for sufficiently high $\text{PSA}_{\text{thr}}$. This simple insight could serve as a valuable guide for enhancing current adaptive chemotherapy treatments. We suggest that clinicians might explore the possibility of implementing HLTC, i.e. waiting for the disease to advance further rather than initiating treatment too early. It is worth mentioning that this kind of ``watchful waiting'' approach before treatment is not uncommon in practice \cite{chapple2002watchful}, at least historically. Furthermore, even without employing the ``watchful waiting'' approach (i.e., disregarding the \textit{high level} component of HLTC), maintaining the current method for determining the appropriate PSA level for the initial drug treatment (that is, $\text{PSA}_\uparrow$) but implementing a \textit{tighter control}, may also change prostate cancer into a chronic disease for many patients (our simulation indicates that, with this approach, 19 out of 32 patients could potentially become chronic).



\section{Discussion}

Cancer has posed an immense challenge to modern medicine due to its complex nature and the absence of a definitive cure \cite{hayden2008cancer}, casting a shadow of devastation upon those affected. This disease can wreak havoc on the human body, resulting in a multitude of severe impacts before leading to the eventual demise. While prostate cancer treatments can prolong life-expectancy, they can also be extremely aggressive and cause debilitating side effects \cite{schirrmacher2019chemotherapy}. Chemotherapy, a common treatment, frequently induces hair loss, nausea, and extreme fatigue, significantly impacting not only the physical but also the emotional well-being of patients \cite{aslam2014side}. It is crucial to understand that augmented drug administration does not always result in an increased lifespan, even at the cost of more suffering. Paradoxically, it appears plausible to significantly extend the patient life while concurrently reducing the treatment ratio (the proportion of time a drug is given compared to the total treatment duration) throughout the treatment course via adaptive chemotherapy \cite{west2019multidrug}. In other words, a longer and better life may be achievable.

In this work, we use a systematic statistical method, LBEB approach for NLME, to estimate patient-informed parameters for a Stackelberg game-theoretic multi-population model that quantitatively describes prostate cancer progression \cite{zhang2022evolution}. We obtain estimates for common parameters across all patients as well as the probability distribution for patient-specific parameters in Section \ref{Extraction}. We then employ a Bayesian Optimization approach to identify the optimal adaptive chemotherapeutic treatment policy for a single drug (Abiraterone) aimed at maximizing the time before the patient begins to show symptoms, under the assumption that we know how advanced the cancer needs to be for symptoms to become manifest, in Section \ref{Optimization}. We show that a \textit{high level tight control} (HLTC) treatment, in which the trigger signals (i.e., the biomarker levels) for drug administration and cessation are both high and close together, typically yields the best outcomes, as demonstrated through both computer-assisted and theoretical means (see Appendix \ref{analytic}). Furthermore, we demonstrate that it may be possible to transform prostate cancer from a terminal disease into a chronic condition for most patients in Section \ref{can_be_chronic}.

{\color{black} An important limitation of our analysis is that we assumed continuous and noiseless PSA monitoring. This is, of course, an idealization. However, this assumption has been addressed partially in our numerical optimization by imposing a lower bound on the control-window size: the on- and off-thresholds are required to differ by at least $10\%$. In other words, the tightest HLTC protocol considered here requires resolving a relative PSA difference of $10\%$. This separation is larger than the reported analytical assay variation for PSA measurements, which is less than $\sim 3\%$ for PSA concentrations below $\sim 20$ng/mL \cite{eastham2003variation}. We also recognize that biological and clinical variability can be larger than analytical assay variability; for example, day-to-day biological variation in total PSA has been reported to have a coefficient of variation of about $\sim 7\%$ \cite{nixon1997biological}. This variability is comparable to, but still below, the $10\%$ control window imposed in our numerical optimization. In addition, with this tightest allowed control window, the typical on/off switching time in our simulations is on the order of a week (e.g. for patients in Fig. 6C and F). This suggests that the monitoring frequency required by the idealized HLTC protocol is not obviously outside clinically feasible timescales. Nevertheless, practical implementation would require appropriate safety margins, repeated measurements, and future optimization procedures that explicitly incorporate discrete sampling and measurement noise.}

Following this work, there are many directions for further exploration. First and foremost is realizing the potential benefits of HLTC adaptive therapeutic treatments in laboratory studies and clinical practice. Work along this direction is already in progress. On the numerical investigation front, we have only {\color{black}optimized} (via Bayesian Optimization) a very simple treatment policy, in which there are only two states to be adjusted in drug administration: either no drug or the maximum dosage rate. Allowing for more structured {\color{black}protocols}, possibly continuous in time -- not just on/off -- as proposed in \cite{cunningham2018optimal}, will provide a much richer protocol space to optimize in, and hopefully, better {\color{black} solutions} to discover. Additional factors can also be considered, such as the extreme toxicity of Docetaxel \cite{west2019multidrug}. Too frequent dosages can be intolerable to the patient's body. Thus, for the treatment policy, limits should be imposed on the administration of Docetaxel within a specific timeframe to ensure patient safety, which is indeed often the case in clinical practices. 


Better analysis on more realistic models of cancer could also lead to new insights and novel discoveries regarding collective behavior emergence from population dynamics under ever-changing conditions \cite{phan2021doesn}. Stochastic therapies (in the spirit of the Kapitza stochastic stabilization of the inverted pendulum \cite{kapitza1951dynamic,sanz2008stabilizing}) might prove more effective in targeting cancerous cells compared to periodic ones (a phenomenon observed in a robotic model system  \cite{wang2022robots}). Importantly, we know that spatial dynamics are crucial in evolutionary systems \cite{wu2013cell}. Considering the metastatic nature of cancer cells, capturing how cells migrate (and also the topology of the environment the cells live in \cite{phan2020bacterial,phan2024social}) in the mathematical models should be of critical importance as we strive for modeling fidelity.

\section{Acknowledgement}

This work was supported by the US National Science Foundation (PHY-1659940 and PHY-1734030), and the Johns Hopkins University Discovery Award 2023-2024.

\appendix

\section{Best Fit Parameters \label{bestfit_params}}

\begin{figure*}[!htbp]
\includegraphics[width=\textwidth]{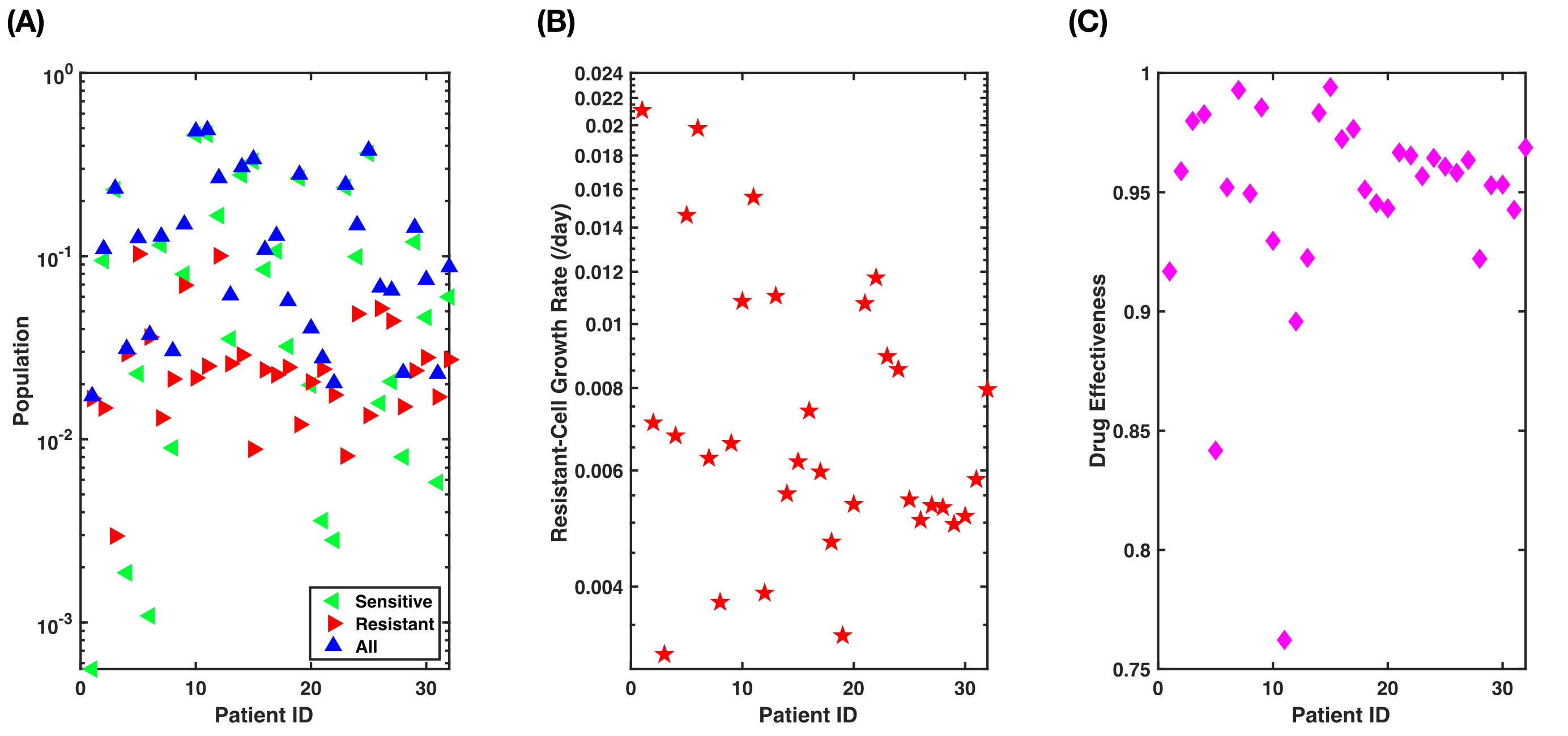}%
\caption{\justifying \textbf{Patient-specific parameters.} Here we show the identified initial population values $\{ x_S(0), x_R(0)\}$, the resistant cell growth rate $r_R$, and the drug effectiveness $\gamma$. \textbf{(A)} The initial values $\{ x_S(0), x_R(0)\}$. The green pointing-left triangles are $x_S(0)$, the red pointing-right triangles are $x_R(0)$, and the blue pointing-up triangles are the sum $x(0)\equiv x_S(0)+x_R(0)$. \textbf{(B)} The resistant cell growth rate $r_R$. \textbf{(C)} The drug effectiveness $\gamma$.}
\label{Patient_Specific}
\end{figure*}

Here, we present the best-fit parameters estimated using our LBEB approach (see Fig. \ref{Patient_Specific}). The patient ID numbers, labeled from 1 to 32, correspond to the following codenames (in the same order): 
\begin{itemize}
    \item Patients treated with adaptive chemotherapy: P1001, P1002, P1003, P1004, P1005, P1006, P1007, P1009, P1010, P1011, P1012, P1014, P1015, P1016, P1017, P1018, P1020.
    \item Patients treated with continuous chemotherapy: C001, C002, C003, C004, C005, C006, C007, C008, C009, C010, C011, C012, C013, C014, C015.
\end{itemize} 

\section{An Analytical Justification for \\ the High Level Tight Control Paradigm \label{analytic}} 

The clinician aims to manage cancer progression such that the observable $\text{PSA}(t)$ remains below a threshold $\text{PSA}_{\text{thr}}$ for as long as possible. This threshold corresponds to keeping the total cancer cell population below $x_{\text{thr}} \equiv \text{PSA}_{\text{thr}}/\text{PSA}_K$. The \textit{total achievable time} $\tau_{tot}$ until this threshold is breached is the sum of three contributions (see Fig. \ref{cartoon}):
\begin{figure*}[!htbp]
\includegraphics[width=\textwidth]{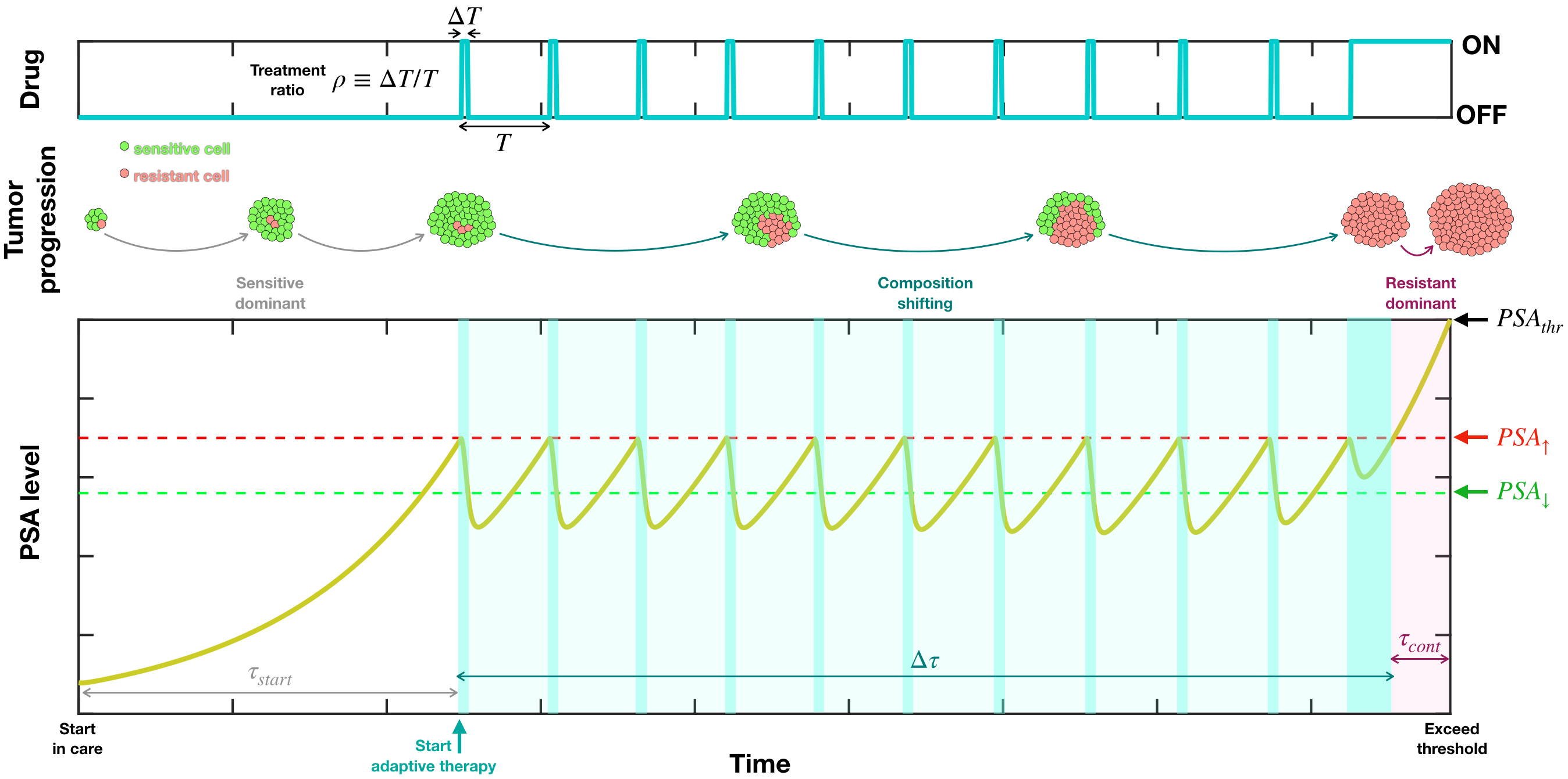}%
\caption{\justifying \textbf{Cancer progression and the changes in PSA levels observed during adaptive chemotherapy treatment.} The upper plot displays drug administration. The middle plot illustrates cancer progression within a tumor, with green circles representing sensitive cells and red circles indicating resistant cells. The bottom plot tracks PSA level changes over time throughout the treatment, highlighting three distinct phases until the threshold is surpassed.}
\label{cartoon}
\end{figure*}
\begin{equation}
    \tau_{\text{tot}} = \tau_{\text{start}} + \Delta \tau + \tau_{\text{cont}} \ ,
\end{equation}
in which we have:
\begin{itemize}
    \item $\tau_{\text{start}}$: The time from when the patient begins care with the clinician (at $t=0$) until the adaptive chemotherapy starts. During this period, the growth of sensitive cells is predominantly observed.
    \item $\Delta \tau$: The drug response time, from when the treatment starts to when it cannot suppress the growth of cancer cells. During this time, the composition of the cancer population shifts from being predominantly sensitive cells to being dominated by resistant cells.
    \item $\tau_{\text{cont}}$: The time from when the treatment fails to when the PSA threshold is breached. During this period, the patient is usually maintained on continuous Abiraterone treatment to prevent any potential regrowth of sensitive cells. Consequently, the growth of resistant cells is predominantly observed. 
\end{itemize}

For a drug treatment cycle, we define the treatment ratio $\rho$ as the fraction of the time period during that cycle in which the drug Abiraterone is administered to the patient. In other words, if the total time period of the treatment cycle is $T$ and the \textit{on-drug} time period is $\Delta T$, then $\rho \equiv \Delta T/T$ (see Fig. \ref{cartoon}).

\subsection{Assumptions and Approximations}

To quantitatively understand why HLTC can typically maximize the \textit{total achievable time} (and also reduce the average Abiraterone intake rate), let us investigate adaptive chemotherapy for prostate cancer in a simple regime to obtain some concrete analytical estimation. We assume that $\alpha_{RS} x_S(t)$ is much larger compared to $x_R(t)$ before Abiraterone first administrated i.e. $t\in [0,\tau_{\text{start}}]$. This condition holds true for most patients at $t=0$ (as shown in Fig. \ref{Any_Pattern}B) and therefore at later time, since the ratio $x_S(t)/x_R(t)$ is always {\color{black}increasing} with time. We can derive that from Eq. \eqref{prostate_cancer_small_model}, as the growth of sensitive cells is always faster: 
\begin{equation}
r_S\left\{1-\left[x_S(t)+x_R(t) \right] \right\} >r_R \left\{1- \left[ x_R(t) + \alpha_{RS} x_S(t) \right] \right\} \ , 
\label{faster_growth}
\end{equation}
in which the ecological competition coefficient $\alpha_{RS}\approx 5.2 \gg 1$ and the growth rate ratio $\beta \equiv r_S/r_R \approx 3.0$ (as estimated in Section \ref{Best_Fit_Est}). We also assume that $\text{PSA}_{\text{thr}}$ is much smaller than $\text{PSA}_K$, which remains reasonable even when even when $\text{PSA}_{\text{thr}} \lesssim \text{PSA}_K/\alpha_{RS}$. Denote the cancer cell population when the biomarker reaches the \textit{on-level} to be $x_\uparrow \equiv \text{PSA}_\uparrow/\text{PSA}_K$, then $x_\uparrow \leq x_\text{thr}$. We therefore have $x_S(t), x_R(t), x_\uparrow \ll 1$ during $t\in [0,\tau_{\text{tot}}]$; however, it is important to note that $\alpha_{RS} x_S(t)$ does not necessarily have to be negligible. These above assumptions allow us to approximate Eq. \eqref{prostate_cancer_small_model} for $t\in [0,\tau_{\text{start}}]$ as:
\begin{equation}
\begin{split}
\frac{d}{dt} x_S(t) \approx r_S x_S(t) & \ ,
\\
\frac{d}{dt} x_R(t) \approx r_R \left[ 1 - \alpha_{RS} x_S(t) \right] x_R(t) & \ .
\end{split}
\label{pop_approx}
\end{equation}
These ODEs can be solved analytically for $x_S(t)$ and $x_R(t)$, given the initial values $x_S(0)$ and $x_R(0)$:
\begin{equation}
    x_S(t) = x_S(0) \exp \left( r_S t \right) \ ,
    \ x_R(t) = x_R(0) \exp\left( r_R t \right) \exp \left\{ -\frac{\alpha_{RS}}{\beta} \left[x_S(t)-x_S(0)\right] \right\} \ .
\label{phase1_sols}
\end{equation}
To obtain $\tau_\text{start}$, we have to solve:
\begin{equation}
    x_S(\tau_{\text{start}}) + x_R(\tau_{\text{start}}) = x_\uparrow \ ,
\end{equation}
which is not analytically solvable in a general case. Let us further assume that the number of sensitive cancer cells dominates when the drug treatment begins i.e. $x_S(\tau_{\text{start}})\gg x_R(\tau_{\text{start}})$, so that $x_S(\tau_{\text{start}}) \approx x_\uparrow$. We can then obtain a simple approximation for $\tau_{\text{start}}$:
\begin{equation}
     \tau_{\text{start}} \approx \ln \left[ \frac{x_\uparrow}{x_S(0)} \right] r_S^{-1} \ .
\label{tau_phase1}
\end{equation}
Plug this back to Eq. \eqref{phase1_sols}, we get the resistant cell population at that time to be:
\begin{equation}
    x_R(\tau_{\text{start}}) = x_R(0) \left[ \frac{x_\uparrow}{x_S(0)} \right]^{1/\beta} \exp \left\{ -\frac{\alpha_{RS}}{\beta} \left[x_\uparrow-x_S(0)\right] \right\} \ .
\end{equation}
If the initial total cancer cell population is small $x(0)\ll x_\uparrow$, then even with $x_S(0)\sim x_R(0)$, the condition $x_S(\tau_{\text{start}}) \sim x_\uparrow \gg x_R(\tau_{\text{start}})$ will still hold. Another way for this condition to arise is if the sensitive cell population initially dominates, meaning $x_S(0)\gg x_R(0)$.

We study an adaptive therapy {\color{black}that} operates between the biomarker levels $\text{PSA}_{\uparrow}$ and $\text{PSA}_{\downarrow} = (1-\epsilon) \text{PSA}_{\uparrow}$. This means the total cancer cell population $x(t)$ is is maintained between $x_\uparrow$ and $x_\uparrow(1-\epsilon)$. Note that from Eq. \eqref{trigger}, we can relate this \textit{control window parameter} with the \textit{off-drug} policy parameter via $\epsilon = 1-\eta_\downarrow$. 

{\color{black} For a general control window $\epsilon$, the adaptive treatment trajectory consists of finite-amplitude nonlinear cycles between $\text{PSA}_{\uparrow}$ and $\text{PSA}_{\downarrow}$, and obtaining a closed-form expression for the total achievable time is difficult. We therefore do not attempt to prove analytically, for arbitrary {\color{black}values} over a general range of $\epsilon$, that the optimum must occur at small $\epsilon$. Instead, we focus on analyzing the small-$\epsilon$ regime, where the dynamics can be averaged over one treatment cycle. This calculation provides a \textit{local analytical} argument showing that, once the system is in this regime, decreasing $\epsilon$ improves the outcome: tighter control decreases the treatment ratio and increases the additional achievable time. Thus, the small-$\epsilon$ (tight control) assumption should be viewed as an analytically tractable limit that explains the \textit{tendency} seen in the full numerical optimization (see Appendix \ref{report}).}

For a tight control $\epsilon \ll 1$, we can assume that during the adaptive treatment time period $t \in [\tau_{\text{start}},\tau_{\text{exit}}]$ (where $\tau_{\text{exit}} \equiv \tau_{\text{start}}+\Delta t$), the cancer population is, \textit{on average}, stabilized at:
\begin{equation}
    x(t) \approx x_\uparrow \left( 1-\frac{\epsilon}2 \right) \ \ \Longrightarrow \ \ x_S(t) \approx x_\uparrow \left( 1-\frac{\epsilon}2 \right)-x_R(t) \ .
\label{stabilize}
\end{equation}
This assumption only holds when $\langle dx/dt \rangle = 0$ is possible, where $\langle \circ \rangle$ denotes the average of a quantity $\circ$ over time during a drug treatment cycle. Since the number of cells always increase during the \textit{off-drug} time period, we need to have during the \textit{on-drug} the following inequality:
\begin{equation}
    \frac{d}{dt} x_S(t)\Big|_{\Lambda(t)=1} + \frac{d}{dt} x_R(t) \leq 0 \ .
\end{equation}
We use Eq. \eqref{prostate_cancer_small_model} to rewrite this as:
\begin{equation}
\begin{split}
     r_S\left\{1-\left[\frac{x_S(t)+x_R(t)}{1-\gamma} \right] \right\} x_S(t) + r_R \left\{1- \left[ x_R(t) + \alpha_{RS} x_S(t)\right] \right\} x_R(t) \leq 0 &
     \\
     \Longrightarrow \ \ \left\{ \left[\frac{1-x_\uparrow \left( 1-\frac{\epsilon}2 \right)}{x_S(t)}\right]-(\alpha_{RS}-1)\right\} x_R(t) \leq \beta\left\{\left[\frac{x_\uparrow \left( 1-\frac{\epsilon}2 \right)}{1-\gamma} \right]-1 \right\} & \ .
\end{split}
\end{equation}
For most patients, $\gamma \rightarrow 1$ (see Fig. \ref{Any_Pattern}A and Fig. \ref{Patient_Specific}C). Together with Eq. \eqref{stabilize} and the assumption $x_S(t), x_R(t), x_\uparrow \ll 1$ previously mentioned, this inequality becomes:
\begin{equation}
      x_S(t) \geq \frac{x_\uparrow}{\beta} \left\{\left[\frac{x_\uparrow \left( 1-\frac{\epsilon}2 \right)}{1-\gamma} \right]-1 \right\}^{-1} \approx \frac{1-\gamma}{\beta \left( 1-\frac{\epsilon}2\right)} \ .
\label{stabilize_fail}
\end{equation}
The right hand side is a finite positive value very close to $0$, which is also increasing with $\epsilon$. This means that the tighter the control (i.e. the smaller the control window), the more difficult it becomes to maintain stabilization Eq. \eqref{stabilize}, as it reduces the allowable range for $x_S$. Therefore, when the adaptive treatment starts failing to suppress the cancer growth (i.e. the inequality does not hold anymore), we can make the following approximation:
\begin{equation}
    x_S(\tau_{\text{exit}})\approx 0 \ \ \Longrightarrow \ \ x_R(\tau_{\text{exit}}) \approx x_\uparrow \left( 1-\frac{\epsilon}2 \right) \ ,
\end{equation}
which is consistent with the behavior seen in Fig. \ref{Adaptive}B,C,E,F and Fig. \ref{Chronic}B. In summary, during the adaptive treatment time period $t \in [\tau_{\text{start}},\tau_{\text{exit}}]$, the growth of Abiraterone resistant cells obeys:
\begin{equation}
\frac{d}{dt} x_R(t) = r_R \left\{ \left[1-\alpha_{RS} x_\uparrow \left( 1-\frac{\epsilon}2 \right)\right] - (1-\alpha_{RS}) x_R(t) \right\} x_R(t) \ ,
\label{resistant_growth_tight}
\end{equation}
which comes from Eq. \eqref{prostate_cancer_small_model} and the stabilization Eq. \eqref{stabilize}, until $x_R(t)$ reaches $x_\uparrow$. The solution of this ODE is given by:
\begin{equation}
    r_R \left[1-\alpha_{RS} x_\uparrow \left( 1-\frac{\epsilon}2 \right)\right] (t-\tau_{\text{start}}) = \ln \left\{ \frac{x_R}{1-\left[ \frac{1-\alpha_{RS}}{1-\alpha_{RS} x_\uparrow \left( 1-\frac{\epsilon}2 \right)} \right]x_R} \right\}\Bigg|^{x_R(t)}_{x_R(\tau_{\text{start})}} \ .
\end{equation}
To obtain $\Delta \tau$ we need to solve:
\begin{equation}
     \Delta \tau = \ln \left\{ \frac{x_R}{1-\left[ \frac{1-\alpha_{RS}}{1-\alpha_{RS} x_\uparrow \left( 1-\frac{\epsilon}2 \right)} \right]x_R} \right\}\Bigg|^{x_\uparrow \left( 1-\frac{\epsilon}2 \right)}_{x_R(\tau_{\text{start})}} \left[\frac{r_R^{-1}}{1-\alpha_{RS} x_\uparrow \left( 1-\frac{\epsilon}2 \right)} \right] \ ,
\end{equation}
in which for $x_R(\tau_\text{start}) \ll x_\uparrow \ll 1$ can help us to further approximate this expression to:
\begin{equation}
    \Delta \tau \approx \ln\left\{ \frac{x_\uparrow \left( 1-\frac{\epsilon}2 \right)\left[1-\alpha_{RS}x_\uparrow  \left( 1-\frac{\epsilon}2 \right) \right]}{x_R(\tau_\text{start})} \right\} \left[\frac{r_R^{-1}}{1-\alpha_{RS} x_\uparrow \left( 1-\frac{\epsilon}2 \right)} \right] \ .
\label{tau_phase2}
\end{equation}

For the last phase $t \in [\tau_{\text{exit}},\tau_{\text{tot}}]$, the resistant cancer cells {\color{black}have} taken over the population, the cancer progression can be described with $x(t)=x_R(t)$ and the ODE:
\begin{equation}
    \frac{d}{dt} x_R(t) \approx r_R \left[ 1- x_R(t) \right] x_R(t) \ ,
\end{equation}
which can be obtained from the Eq. \eqref{prostate_cancer_small_model} after setting $x_S(t)=0$. This ODE can be solved analytically to give:
\begin{equation}
    r_R (t-\tau_{\text{exit}}) = \ln \left[ \frac{x_R}{1-x_R} \right]\Bigg|^{x_R(t)}_{x_R(\tau_{\text{exit})}} \ .
\end{equation}
To obtain $\tau_\text{cont}$ we need to solve:
\begin{equation}
    \tau_\text{cont} = \ln \left[ \frac{x_R}{1-x_R} \right]\Bigg|^{x_\text{thr}}_{x_\uparrow \left( 1-\frac{\epsilon}2\right)} r_R^{-1} \ ,
\end{equation}
in which we can use $x_\text{thr}\ll 1$ to to approximate its value to be:
\begin{equation}
    \tau_\text{cont} = \ln \left[ \frac{x_\text{thr}}{x_\uparrow \left( 1-\frac{\epsilon}2\right)} \right] r_R^{-1} \ .
\label{tau_phase3}
\end{equation}

\subsection{An Analytical Expression for the Treatment Ratio $\rho$ \\ and the Total Achievable Time $\tau_{\text{tot}}$}

\subsubsection{The Treatment Ratio $\rho$}

Define the following averaging during a drug treatment cycle:
\begin{equation}
\begin{split}
\Omega \equiv \left\langle \frac{1}{1-\gamma \Lambda (t)}\right\rangle &= \rho \times \left(\frac{1}{1-\gamma}\right)+ (1-\rho) \times 1  
\\
&= 1 + \left(\frac{\gamma}{1-\gamma}\right)\rho \ \Longrightarrow \ \rho = \left(\frac{1-\gamma}{\gamma}\right) (\Omega-1)  \ , 
\end{split}
\end{equation}
so that when Eq. \eqref{stabilize} holds, we can use Eq. \eqref{prostate_cancer_small_model} to obtain:
\begin{equation}
\begin{split}
    \frac{d}{dt}\left[ x_S + x_R \right] = r_S \left[ 1-\Omega \Big( x_S + x_R \Big) \right] x_S + r_R \left[ 1 - \Big(x_R+\alpha_{RS} x_S \Big)\right] x_R = 0 &
    \\
    \Longrightarrow \ \rho = \left(\frac{1-\gamma}{\gamma}\right) \left( \left\{ \frac{1-\frac{x_R}{\beta x_S}\left[ 1 - \Big(x_R+\alpha_{RS} x_S \Big)\right]}{x_\uparrow \left( 1-\frac{\epsilon}2\right)} \right\} -1 \right) \ & \ .
\end{split}
\label{gen_ratio}
\end{equation}
If the resistant cell population is negligible i.e. $x_R \ll x_S$ (and note that $\gamma \rightarrow 1$ for most patients) then this expression can be further simplified:
\begin{equation}
    \rho \approx \frac{1-\gamma}{x_\uparrow \left( 1-\frac{\epsilon}2\right)} \ .
\end{equation}
This implies the tightest control window $\epsilon$ (thus the maximum policy parameter $\eta_\downarrow$) will give the smallest treatment ratio $\rho$.

\subsubsection{The Total Achievable Time $\tau_{\text{tot}}$}

The \textit{total achievable time} can be calculated from the sum of the times given in Eq. \eqref{tau_phase1}, Eq. \eqref{tau_phase2}, and Eq. \eqref{tau_phase3}; however, the resulting expression is quite complex. In the limit of very {\color{black} few} resistant cells in the beginning $x_R(0)\rightarrow 0$, the formula for $\tau_{\text{tot}}$ is simple and can be {\color{black}split} into two parts:
\begin{equation}
    \tau_{\text{tot}} = \tau_{\text{tot}}^{\text{(c)}} + \tau_{\text{tot}}^{\text{(a)}} \ .
\end{equation}
The part $\tau^{(c)}_{\text{tot}}$ is equal to the \textit{total achievable time} if a continuous drug treatment is applied from the beginning, while $\tau^{(a)}_{\text{tot}}$ represents the additional time gained by using an adaptive chemotherapeutic treatment (for a given control window $\epsilon$ and \textit{on}-drug level $\text{PSA}_\uparrow$) instead:
\begin{equation}
\begin{split}
    \tau_{\text{tot}}^{\text{(c)}} \approx \ln\left[ \frac{x_{\text{thr}}}{(1-x_{\text{thr}}) x_R(0)}\right] r_R^{-1} & \ ,
    \\
    \tau_{\text{tot}}^{\text{(a)}} \approx \left[ \frac{\alpha_{RS} x_\uparrow \left(1-\frac{\epsilon}2\right)}{1-\alpha_{RS} x_\uparrow\left(1-\frac{\epsilon}2\right)} \right] \ln \left\{ \frac{x_\uparrow}{\left[1-x_\uparrow\left(1-\frac{\epsilon}2\right)\right] x_R(0)} \right\} r_R^{-1} & \ .
\end{split}
\label{split_two_tau_tot}
\end{equation}
Consider increasing the $x_\text{thr}$ {\color{black} without surpassing} $ 1/\alpha_{RS}$, then maximum time gained corresponds to the \textit{tightest control} (i.e. smallest possible $\epsilon$) and the \textit{highest} PSA \textit{on}-drug level below the threshold (i.e. largest allowable $x_\uparrow$):
\begin{equation}
    \max \tau_{\text{tot}}^{\text{(a)}} \leq  \tau_{\text{tot}}^{\text{(a)}}(0,x_\text{thr}) =  \left( \frac{\alpha_{RS} x_\text{thr}}{1-\alpha_{RS} x_\text{thr}} \right) \tau_{\text{tot}}^{\text{(c)}} \ .
\end{equation}
This description is that of the HLTC treatment. If we let $x_\text{thr}$ exceed $1/\alpha_{RS}$, then $\max \tau_{\text{tot}}^{\text{(a)}}$ can be infinitely large, indicating that the cancer disease can become chronic. Note that for this to occur, the total number of cancer cells must be stabilized, as indicated in Eq. \eqref{stabilize}. This stabilization can fail if the population of sensitive cells is too low, as explained in Eq. \eqref{stabilize_fail}.

We have made several assumptions and approximations to reach the conclusions outlined above. While our findings may not be universally applicable, they can serve as a valuable guide, as they {\color{black} agree} with the optimal treatment strategy for most cases (41 out of 43) discussed in Appendix \ref{report}.

\section{Optimal Treatment Policy Report \label{report}}

We report the optimal adaptive therapy treatment policy parameters $(\eta_\uparrow^*,\eta_\downarrow^*)$ found by employing a Bayesian optimization approach, with different values of the PSA threshold i.e. $\text{PSA}_\text{thr}/\text{PSA}_K=0.14$ in Table \ref{table1} and $\text{PSA}_\text{thr}/\text{PSA}_K=0.19$ in Table \ref{table2}. After the results found in Section \ref{optim}, instead of {\color{black}selecting} $n_{\text{eval}}=10$ random points to evaluate at the beginning, we do that for $n_{\text{eval}}-2=8$ and let the last two points be $(\eta_\uparrow,\eta_\downarrow)=(0.00,0.00)$ (representing a continuous treatment start at $t=0$) and $(\eta_\uparrow,\eta_\downarrow)=(0.90,0.90)$ (representing a HLTC treatment).
      
\begin{table}
\centering
\begin{tabular}{ |c|c|c|c|c|c| } 
\hline
Patient ID & $\eta^*_\uparrow$ & $\eta^*_\downarrow$ & $\tau_{\text{tot}}(\eta^*_\uparrow,\eta^*_\uparrow)$ & $\tau_{\text{tot}}(0.90,0.90)$ & $\tau_{\text{tot}}(0.00,0.00)$ \\
\hline
P1001 & 0.90 & 9.00 & 112 & 112 & 108 \\ 
P1002 & 0.90 & 0.90 & 544 & 544 & 376 \\ 
P1004 & 0.90 & 0.90 & 271 & 271 & 249 \\
P1005 & 0.22 & 0.81 & 10 & 10 & 10 \\
P1006 & 0.90 & 0.90 & 76 & 76 & 74 \\
P1007 & 0.90 & 0.90 & 664 & 664 & 406 \\
P1009 & 0.90 & 0.90 & 680 & 680 & 555 \\
P1015 & 0.90 & 0.90 & 223 & 223 & 189 \\
P1018 & 0.90 & 0.90 & 382 & 382 & 271 \\
P1020 & 0.90 & 0.90 & 495 & 495 & 345 \\
C001 & 0.90 & 0.90 & 556 & 556 & 431 \\
C003 & 0.90 & 0.90 & 531 & 531 & 418 \\
C004 & 0.90 & 0.90 & 205 & 205 & 177 \\
C005 & 0.90 & 0.90 & 228 & 228 & 191 \\
C009 & 0.90 & 0.90 & 248 & 248 & 223 \\
C010 & 0.90 & 0.90 & 285 & 285 & 244 \\
C011 & 0.90 & 0.90 & 593 & 593 & 495 \\
C013 & 0.90 & 0.90 & 474 & 474 & 370 \\
C014 & 0.90 & 0.90 & 494 & 494 & 405 \\
C015 & 0.90 & 0.90 & 317 & 317 & 234 \\
\hline
\end{tabular}
\caption{$\text{PSA}_\text{thr}/\text{PSA}_K=0.14$}
\label{table1}
\end{table}

\begin{table}
\centering
\begin{tabular}{|c|c|c|c|c|c|}
\hline
Patient ID & $\eta^*_\uparrow$ & $\eta^*_\downarrow$ & $\tau_{\text{tot}}(\eta^*_\uparrow,\eta^*_\uparrow)$ & $\tau_{\text{tot}}(0.90,0.90)$ & $\tau_{\text{tot}}(0.00,0.00)$ \\
\hline
P1001 & 0.90 & 0.90 & 147 & 147 & 126 \\ 
P1002 & 0.90 & 0.90 & 959 & 959 & 427 \\ 
P1004 & 0.90 & 0.90 & 358 & 358 & 303 \\
P1005 & 0.41 & 0.68 & 45 & 38 & 45 \\
P1006 & 0.90 & 0.90 & 101 & 101 & 93 \\
P1007 & 0.90 & 0.90 & 1196 & 1196 & 464 \\
P1009 & 0.90 & 0.90 & 1005 & 1005 & 652 \\
P1010 & 0.90 & 0.90 & 244 & 244 & 177\\
P1015 & 0.90 & 0.90 & 365 & 365 & 230 \\
P1018 & 0.90 & 0.90 & 646 & 646 & 320 \\
P1020 & 0.90 & 0.90 & 853 & 853 & 406 \\
C001 & 0.90 & 0.90 & 884 & 884 & 511 \\
C003 & 0.90 & 0.90 & 831 & 831 & 489 \\
C004 & 0.90 & 0.90 & 282 & 282 & 211 \\
C005 & 0.90 & 0.90 & 316 & 316 & 221 \\
C007 & 0.90 & 0.90 & 295 & 295 & 191 \\
C009 & 0.90 & 0.90 & 377 & 377 & 296 \\
C010 & 0.90 & 0.90 & 435 & 435 & 313 \\
C011 & 0.90 & 0.90 & 913 & 913 & 579 \\
C012 & 0.90 & 0.90 & 985 & 985 & 511 \\
C013 & 0.90 & 0.90 & 769 & 769 & 442 \\
C014 & 0.90 & 0.90 & 724 & 724 & 470 \\
C015 & 0.90 & 0.90 & 522 & 522 & 280 \\
\hline
\end{tabular}
\caption{$\text{PSA}_\text{thr}/\text{PSA}_K=0.19$}
\label{table2}
\end{table}

\begin{figure*}[!htbp]
\includegraphics[width=\textwidth]{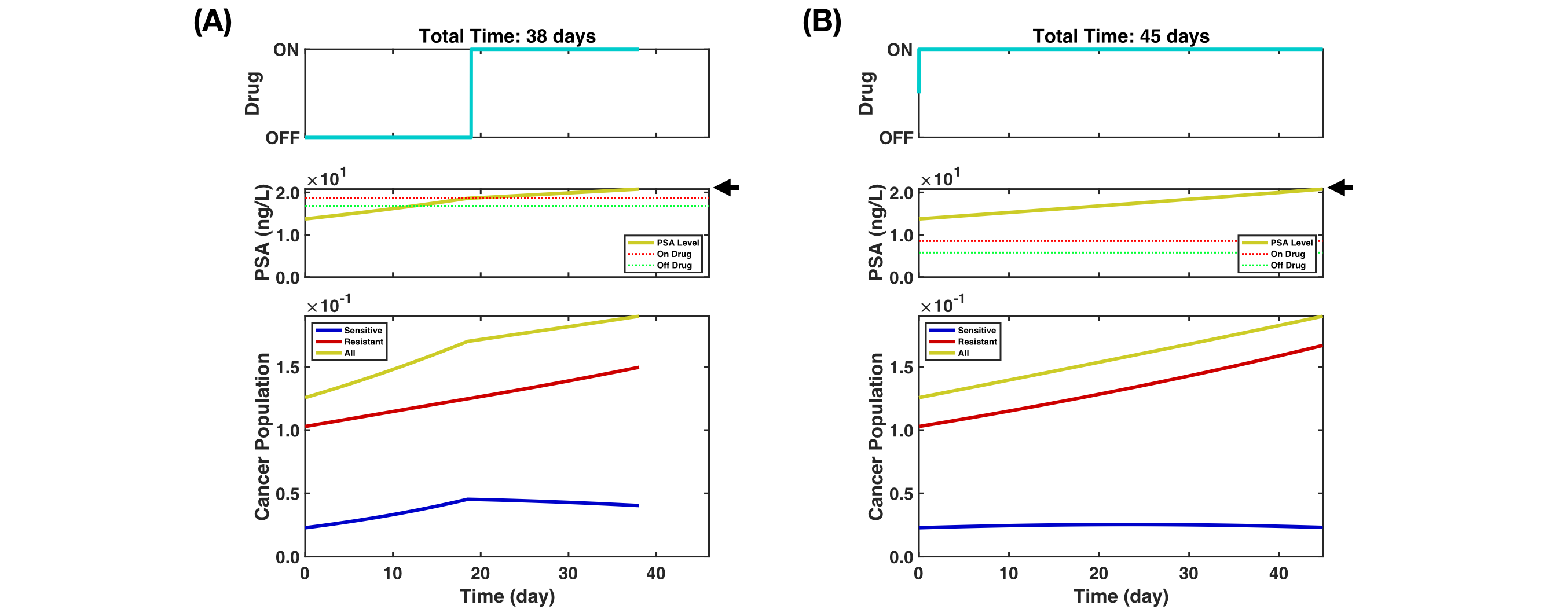}%
\caption{\justifying \textbf{Cancer progression of patient P1005 under the PSA threshold $\text{PSA}_\text{thr}/\text{PSA}_K=0.19$.} This patient has a low drug effectiveness {\color{black}parameter}, so that when the drug is administrated early, it only has a small impact. We show the response of this patient
P1007 under the HLTC adaptive therapy with $(\eta_\uparrow,\eta_\downarrow)=(0.90,0.90)$ in \textbf{(A)}  \textbf{(A)}, and under the optimal adaptive therapy (identified via Bayesian optimization) with $(\eta^*_\uparrow,\eta^*_\downarrow)=(0.41,0.68)$ in \textbf{(B)}, which is equivalent to a continuous treatment started from the very beginning. The black arrow indicates the PSA threshold for each plot.}
\label{P1005}
\end{figure*}

There is only one patient that has $(\eta^*_\uparrow,\eta^*_\downarrow) \neq (0.90,0.90)$ for both of these $\text{PSA}_{\text{thr}}/\text{PSA}_K$, which is patient P1005. Upon closer examination, we find that this patient has very low drug effectiveness, indicating that for the PSA threshold values we investigated, it is preferable to initiate treatment as soon as possible. In Fig. \ref{P1005}, we show the PSA progressions under $\text{PSA}_{\text{thr}}/\text{PSA}_K=0.19$, for the adaptive treatment using HLTC $(\eta_\uparrow,\eta_\downarrow)=(0.90,0.90)$ and for the optimal treatment $(\eta_\uparrow^*,\eta_\downarrow^*)$ identified by our Bayesian optimization approach. Note that $\text{PSA}_\uparrow$ for the found optimal treatment is already below $\text{PSA}(0)$, and $\text{PSA}(t)$ does not decrease, indicating that it is effectively equivalent to a continuous treatment initiated from the very beginning at $t=0$.  

\ \ 

\bibliography{main}

@article{schwarz1978estimating,
  title={Estimating the dimension of a model},
  author={Schwarz, Gideon},
  journal={The annals of statistics},
  pages={461--464},
  year={1978},
  publisher={JSTOR}
}

@article{akaike1974new,
  title={A new look at the statistical model identification},
  author={Akaike, Hirotugu},
  journal={IEEE transactions on automatic control},
  volume={19},
  number={6},
  pages={716--723},
  year={1974},
  publisher={Ieee}
}

@article{eastham2003variation,
  title={Variation of serum prostate-specific antigen levels: an evaluation of year-to-year fluctuations},
  author={Eastham, James A and Riedel, Elyn and Scardino, Peter T and Shike, Moshe and Fleisher, Martin and Schatzkin, Arthur and Lanza, Elaine and Latkany, Lianne and Begg, Colin B and Polyp Prevention Trial Study Group and others},
  journal={Jama},
  volume={289},
  number={20},
  pages={2695--2700},
  year={2003},
  publisher={American Medical Association}
}

@article{nixon1997biological,
  title={Biological variation of prostate specific antigen levels in serum: an evaluation of day-to-day physiological fluctuations in a well-defined cohort of 24 patients},
  author={Nixon, Randy G and Wener, Mark H and Smith, Katie M and Parson, Robert E and Strobel, Susan A and Brawer, Michael K},
  journal={The Journal of urology},
  volume={157},
  number={6},
  pages={2183--2190},
  year={1997},
  publisher={Wolters Kluwer Philadelphia, PA}
}

@article{leuprorelin_action,
  title={Leuprorelin (Leuplin, Lupron, Viadur) A Prostate Cancer Drug with Dual Innovations in Mechanism of Action and Drug Delivery System},
  author={Takada, Naoki and Kawabe, Hideo},
  journal={Drug Discovery in Japan: Investigating the Sources of Innovation},
  pages={65--84},
  year={2019},
  publisher={Springer}
}

@article{resource_game_theory,
  title={Game Theory for Managing Evolving Systems: Challenges and Opportunities of Including Vector-Valued Strategies and Life-History Traits},
  author={Kleshnina, Maria and Streipert, Sabrina and Brown, Joel S and Sta{\v{n}}kov{\'a}, Kate{\v{r}}ina},
  journal={Dynamic Games and Applications},
  volume={13},
  number={4},
  pages={1130--1155},
  year={2023},
  publisher={Springer}
}

@article{adrenal_production,
  title={Androgen synthesis inhibitors in the treatment of castration-resistant prostate cancer},
  author={Stein, Mark N and Patel, Neal and Bershadskiy, Alexander and Sokoloff, Alisa and Singer, Eric A},
  journal={Asian journal of andrology},
  volume={16},
  number={3},
  pages={387--400},
  year={2014},
  publisher={Medknow}
}

@article{leuprolide_resistance,
  title={Androgen deprivation therapy with Leuprolide acetate for treatment of advanced prostate cancer},
  author={Hoda, M Raschid and Kramer, Mario W and Merseburger, Axel S and Cronauer, Marcus V},
  journal={Expert opinion on pharmacotherapy},
  volume={18},
  number={1},
  pages={105--113},
  year={2017},
  publisher={Taylor \& Francis}
}

@article{androgen_sensitive,
  title={Metastatic prostate cancer—a review of current treatment options and promising new approaches},
  author={Posdzich, Philip and Darr, Christopher and Hilser, Thomas and Wahl, Milan and Herrmann, Ken and Hadaschik, Boris and Gr{\"u}nwald, Viktor},
  journal={Cancers},
  volume={15},
  number={2},
  pages={461},
  year={2023},
  publisher={MDPI}
}

@article{deprivation,
  title={Androgen Signaling in Prostate Cancer: When a Friend Turns Foe.},
  author={Pandey, Swaroop Kumar and Sabharwal, Usha and Tripathi, Swati and Mishra, Anuja and Yadav, Neha and Dwivedi-Agnihotri, Hemlata},
  journal={Endocrine, Metabolic \& Immune Disorders Drug Targets},
  year={2024}
}

@article{androgens,
  title={Decoding androgen receptor signalling: Genomic vs. non-genomic roles in prostate cancer},
  author={Asim, Mohammad},
  journal={Neoplasia},
  volume={58},
  pages={101066},
  year={2024},
  publisher={Elsevier}
}

@article{oesterling1993effect,
  title={Effect of cystoscopy, prostate biopsy, and transurethral resection of prostate on serum prostate-specific antigen concentration},
  author={Oesterling, Joseph E and Rice, David C and Glenski, William J and Bergstralh, Erik J},
  journal={Urology},
  volume={42},
  number={3},
  pages={276--282},
  year={1993},
  publisher={Elsevier}
}

@article{fizazi2017abiraterone,
  title={Abiraterone plus prednisone in metastatic, castration-sensitive prostate cancer},
  author={Fizazi, Karim and Tran, NamPhuong and Fein, Luis and Matsubara, Nobuaki and Rodriguez-Antolin, Alfredo and Alekseev, Boris Y and {\"O}zg{\"u}ro{\u{g}}lu, Mustafa and Ye, Dingwei and Feyerabend, Susan and Protheroe, Andrew and others},
  journal={New England Journal of Medicine},
  volume={377},
  number={4},
  pages={352--360},
  year={2017},
  publisher={Mass Medical Soc}
}

@inproceedings{hashimoto2019serum,
  title={Serum testosterone level is a useful biomarker for determining the optimal treatment for castration-resistant prostate cancer},
  author={Hashimoto, Kohei and Tabata, Hidetoshi and Shindo, Tetsuya and Tanaka, Toshiaki and Hashimoto, Jiro and Inoue, Ryuta and Muranaka, Takashi and Hotta, Hiroshi and Yanase, Masahiro and Kunishima, Yasuharu and others},
  booktitle={Urologic Oncology: Seminars and Original Investigations},
  volume={37},
  pages={485--491},
  year={2019},
  organization={Elsevier}
}

@article{zhang_2017,
author = {Zhang, J. S. and Cunningham, J. J. and Brown, J. S. and Gatenby, R. A.},
   title = {Integrating evolutionary dynamics into treatment of metastatic castrate-resistant prostate cancer},
   journal = {Nature Communications},
   volume = {8},
   note = {Fn9yp
Times Cited:235
Cited References Count:36},
   year = {2017}
}

@article{tumor-micro,
author = {Ge, R. B. and Wang, Z. W. and Cheng, L.},
   title = {Tumor microenvironment heterogeneity an important mediator of prostate cancer progression and therapeutic resistance},
   journal = {Npj Precision Oncology},
   volume = {6},
   number = {1},
   note = {0z0sc
Times Cited:50
Cited References Count:138},
   year = {2022}
}

@article{stackelberg,
  title={Stackelberg evolutionary game theory: how to manage evolving systems},
  author={Stein, Alexander and Salvioli, Monica and Garjani, Hasti and Dubbeldam, Johan and Viossat, Yannick and Brown, Joel S and Sta{\v{n}}kov{\'a}, Kate{\v{r}}ina},
  journal={Philosophical Transactions of the Royal Society B},
  volume={378},
  number={1876},
  pages={20210495},
  year={2023},
  publisher={The Royal Society}
}

@article{hansen-read,
  title={Modifying adaptive therapy to enhance competitive suppression},
  author={Hansen, Elsa and Read, Andrew F},
  journal={Cancers},
  volume={12},
  number={12},
  pages={3556},
  year={2020},
  publisher={MDPI}
}

@article{efron1978regression,
  title={Regression and ANOVA with zero-one data: Measures of residual variation},
  author={Efron, Bradley},
  journal={Journal of the American Statistical Association},
  volume={73},
  number={361},
  pages={113--121},
  year={1978},
  publisher={Taylor \& Francis}
}

@book{schabenberger2001contemporary,
  title={Contemporary statistical models for the plant and soil sciences},
  author={Schabenberger, Oliver and Pierce, Francis J},
  year={2001},
  publisher={CRC press}
}

@article{spiess2010evaluation,
  title={An evaluation of R 2 as an inadequate measure for nonlinear models in pharmacological and biochemical research: a Monte Carlo approach},
  author={Spiess, Andrej-Nikolai and Neumeyer, Natalie},
  journal={BMC pharmacology},
  volume={10},
  pages={1--11},
  year={2010},
  publisher={Springer}
}

@misc{mcfadden1972conditional,
  title={Conditional logit analysis of qualitative choice behavior},
  author={McFadden, Daniel},
  year={1972}
}

@book{montgomery2021introduction,
  title={Introduction to linear regression analysis},
  author={Montgomery, Douglas C and Peck, Elizabeth A and Vining, G Geoffrey},
  year={2021},
  publisher={John Wiley \& Sons}
}

@book{crow1987lognormal,
  title={Lognormal distributions},
  author={Crow, Edwin L and Shimizu, Kunio},
  year={1987},
  publisher={Marcel Dekker New York}
}

@article{lindstrom1990nonlinear,
  title={Nonlinear mixed effects models for repeated measures data},
  author={Lindstrom, Mary J and Bates, Douglas M},
  journal={Biometrics},
  pages={673--687},
  year={1990},
  publisher={JSTOR}
}

@article{zhang2022evolution,
  title={Evolution-based mathematical models significantly prolong response to abiraterone in metastatic castrate-resistant prostate cancer and identify strategies to further improve outcomes},
  author={Zhang, Jingsong and Cunningham, Jessica and Brown, Joel and Gatenby, Robert},
  journal={Elife},
  volume={11},
  pages={e76284},
  year={2022},
  publisher={eLife Sciences Publications Limited}
}

@article{anderson1994model,
  title={Model comparisons and R 2},
  author={Anderson-Sprecher, Richard},
  journal={The American Statistician},
  volume={48},
  number={2},
  pages={113--117},
  year={1994},
  publisher={Taylor \& Francis}
}

@article{smith2013comparison,
  title={A comparison of logistic regression pseudo R2 indices},
  author={Smith, Thomas J and McKenna, Cornelius M},
  journal={Multiple Linear Regression Viewpoints},
  volume={39},
  number={2},
  pages={17--26},
  year={2013},
  publisher={Akron}
}

@article{sottoriva2015big,
  title={A Big Bang model of human colorectal tumor growth},
  author={Sottoriva, Andrea and Kang, Haeyoun and Ma, Zhicheng and Graham, Trevor A and Salomon, Matthew P and Zhao, Junsong and Marjoram, Paul and Siegmund, Kimberly and Press, Michael F and Shibata, Darryl and others},
  journal={Nature genetics},
  volume={47},
  number={3},
  pages={209--216},
  year={2015},
  publisher={Nature Publishing Group}
}

@article{bhang2015studying,
  title={Studying clonal dynamics in response to cancer therapy using high-complexity barcoding},
  author={Bhang, Hyo-eun C and Ruddy, David A and Krishnamurthy Radhakrishna, Viveksagar and Caushi, Justina X and Zhao, Rui and Hims, Matthew M and Singh, Angad P and Kao, Iris and Rakiec, Daniel and Shaw, Pamela and others},
  journal={Nature medicine},
  volume={21},
  number={5},
  pages={440--448},
  year={2015},
  publisher={Nature Publishing Group US New York}
}

@article{turke2010preexistence,
  title={Preexistence and clonal selection of MET amplification in EGFR mutant NSCLC},
  author={Turke, Alexa B and Zejnullahu, Kreshnik and Wu, Yi-Long and Song, Youngchul and Dias-Santagata, Dora and Lifshits, Eugene and Toschi, Luca and Rogers, Andrew and Mok, Tony and Sequist, Lecia and others},
  journal={Cancer cell},
  volume={17},
  number={1},
  pages={77--88},
  year={2010},
  publisher={Elsevier}
}

@article{ramirez2016diverse,
  title={Diverse drug-resistance mechanisms can emerge from drug-tolerant cancer persister cells},
  author={Ramirez, Michael and Rajaram, Satwik and Steininger, Robert J and Osipchuk, Daria and Roth, Maike A and Morinishi, Leanna S and Evans, Louise and Ji, Weiyue and Hsu, Chien-Hsiang and Thurley, Kevin and others},
  journal={Nature communications},
  volume={7},
  number={1},
  pages={10690},
  year={2016},
  publisher={Nature Publishing Group UK London}
}

@article{hata2016tumor,
  title={Tumor cells can follow distinct evolutionary paths to become resistant to epidermal growth factor receptor inhibition},
  author={Hata, Aaron N and Niederst, Matthew J and Archibald, Hannah L and Gomez-Caraballo, Maria and Siddiqui, Faria M and Mulvey, Hillary E and Maruvka, Yosef E and Ji, Fei and Bhang, Hyo-eun C and Krishnamurthy Radhakrishna, Viveksagar and others},
  journal={Nature medicine},
  volume={22},
  number={3},
  pages={262--269},
  year={2016},
  publisher={Nature Publishing Group US New York}
}

@article{hoffmann2000environmental,
  title={Environmental stress as an evolutionary force},
  author={Hoffmann, Ary A and Hercus, Miriam J},
  journal={Bioscience},
  volume={50},
  number={3},
  pages={217--226},
  year={2000},
  publisher={American Institute of Biological Sciences}
}

@article{fitzgerald2017stress,
  title={Stress-induced mutagenesis: implications in cancer and drug resistance},
  author={Fitzgerald, Devon M and Hastings, PJ and Rosenberg, Susan M},
  journal={Annual Review of Cancer Biology},
  volume={1},
  pages={119--140},
  year={2017},
  publisher={Annual Reviews}
}

@article{gatenby2009adaptive,
  title={Adaptive therapy},
  author={Gatenby, Robert A and Silva, Ariosto S and Gillies, Robert J and Frieden, B Roy},
  journal={Cancer research},
  volume={69},
  number={11},
  pages={4894--4903},
  year={2009},
  publisher={AACR}
}

@article{de2011abiraterone,
  title={Abiraterone and increased survival in metastatic prostate cancer},
  author={De Bono, Johann S and Logothetis, Christopher J and Molina, Arturo and Fizazi, Karim and North, Scott and Chu, Luis and Chi, Kim N and Jones, Robert J and Goodman Jr, Oscar B and Saad, Fred and others},
  journal={New England Journal of Medicine},
  volume={364},
  number={21},
  pages={1995--2005},
  year={2011},
  publisher={Mass Medical Soc}
}

@article{kapitza1951dynamic,
  title={Dynamic stability of a pendulum with an oscillating point of support},
  author={Kapitza, PL},
  journal={Zh. Eksp. Teor. Fiz},
  volume={21},
  pages={588},
  year={1951}
}

@article{sanz2008stabilizing,
  title={STABILIZING WITH A HAMMER.},
  author={Sanz-Serna, JM},
  journal={Stochastics \& Dynamics},
  volume={8},
  number={1},
  year={2008}
}

@article{phan2024social,
  title={Social Physics of Bacteria: Avoidance of an Information Black Hole},
  author={Phan, Trung V and Li, Shengkai and Ferreris, Domenic and Morris, Ryan and Bos, Julia and Gou, Buming and Martiniani, Stephano and Chaikin, Paul and Kevrekidis, Yannis G and Austin, Robert H},
  journal={arXiv preprint arXiv:2401.16691},
  year={2024}
}

@article{west2019multidrug,
  title={Multidrug cancer therapy in metastatic castrate-resistant prostate cancer: an evolution-based strategy},
  author={West, Jeffrey B and Dinh, Mina N and Brown, Joel S and Zhang, Jingsong and Anderson, Alexander R and Gatenby, Robert A},
  journal={Clinical Cancer Research},
  volume={25},
  number={14},
  pages={4413--4421},
  year={2019},
  publisher={AACR}
}

@article{west2020towards,
  title={Towards multidrug adaptive therapy},
  author={West, Jeffrey and You, Li and Zhang, Jingsong and Gatenby, Robert A and Brown, Joel S and Newton, Paul K and Anderson, Alexander RA},
  journal={Cancer research},
  volume={80},
  number={7},
  pages={1578--1589},
  year={2020},
  publisher={AACR}
}

@article{oliver1990kriging,
  title={Kriging: a method of interpolation for geographical information systems},
  author={Oliver, Margaret A and Webster, Richard},
  journal={International Journal of Geographical Information System},
  volume={4},
  number={3},
  pages={313--332},
  year={1990},
  publisher={Taylor \& Francis}
}

@article{phan2021doesn,
  title={It doesn’t always pay to be fit: success landscapes},
  author={Phan, Trung V and Wang, Gao and Do, Tuan K and Kevrekidis, Ioannis G and Amend, Sarah and Hammarlund, Emma and Pienta, Ken and Brown, Joel and Liu, Liyu and Austin, Robert H},
  journal={Journal of Biological Physics},
  volume={47},
  pages={387--400},
  year={2021},
  publisher={Springer}
}

@article{frazier2018tutorial,
  title={A tutorial on Bayesian optimization},
  author={Frazier, Peter I},
  journal={arXiv preprint arXiv:1807.02811},
  year={2018}
}

@article{schirrmacher2019chemotherapy,
  title={From chemotherapy to biological therapy: A review of novel concepts to reduce the side effects of systemic cancer treatment},
  author={Schirrmacher, Volker},
  journal={International journal of oncology},
  volume={54},
  number={2},
  pages={407--419},
  year={2019},
  publisher={Spandidos Publications}
}

@article{aslam2014side,
  title={Side effects of chemotherapy in cancer patients and evaluation of patients opinion about starvation based differential chemotherapy},
  author={Aslam, Muhammad Shahbaz and Naveed, Sidra and Ahmed, Aftab and Abbas, Zaigham and Gull, Iram and Athar, Muhammad Amin},
  journal={Journal of Cancer Therapy},
  volume={2014},
  year={2014},
  publisher={Scientific Research Publishing}
}

@article{phan2020bacterial,
  title={Bacterial route finding and collective escape in mazes and fractals},
  author={Phan, Trung V and Morris, Ryan and Black, Matthew E and Do, Tuan K and Lin, Ke-Chih and Nagy, Krisztina and Sturm, James C and Bos, Julia and Austin, Robert H},
  journal={Physical Review X},
  volume={10},
  number={3},
  pages={031017},
  year={2020},
  publisher={APS}
}

@article{wu2013cell,
  title={Cell motility and drug gradients in the emergence of resistance to chemotherapy},
  author={Wu, Amy and Loutherback, Kevin and Lambert, Guillaume and Est{\'e}vez-Salmer{\'o}n, Luis and Tlsty, Thea D and Austin, Robert H and Sturm, James C},
  journal={Proceedings of the National Academy of Sciences},
  volume={110},
  number={40},
  pages={16103--16108},
  year={2013},
  publisher={National Acad Sciences}
}

@article{cunningham2018optimal,
  title={Optimal control to develop therapeutic strategies for metastatic castrate resistant prostate cancer},
  author={Cunningham, Jessica J and Brown, Joel S and Gatenby, Robert A and Sta{\v{n}}kov{\'a}, Kate{\v{r}}ina},
  journal={Journal of theoretical biology},
  volume={459},
  pages={67--78},
  year={2018},
  publisher={Elsevier}
}

@article{hayden2008cancer,
  title={Cancer complexity slows quest for cure},
  author={Hayden, E Check and others},
  journal={Nature},
  volume={455},
  number={7210},
  pages={148},
  year={2008},
  publisher={Nature}
}

@article{schulz2018tutorial,
  title={A tutorial on Gaussian process regression: Modelling, exploring, and exploiting functions},
  author={Schulz, Eric and Speekenbrink, Maarten and Krause, Andreas},
  journal={Journal of Mathematical Psychology},
  volume={85},
  pages={1--16},
  year={2018},
  publisher={Elsevier}
}

@article{wang2022robots,
  title={Robots as models of evolving systems},
  author={Wang, Gao and Phan, Trung V and Li, Shengkai and Wang, Jing and Peng, Yan and Chen, Guo and Qu, Junle and Goldman, Daniel I and Levin, Simon A and Pienta, Kenneth and others},
  journal={Proceedings of the National Academy of Sciences},
  volume={119},
  number={12},
  pages={e2120019119},
  year={2022},
  publisher={National Acad Sciences}
}

@article{chapple2002watchful,
  title={Is ‘watchful waiting’a real choice for men with prostate cancer? A qualitative study},
  author={Chapple, A and Ziebland, S and Herxheimer, A and McPherson, A and Shepperd, S and Miller, R},
  journal={BJU international},
  volume={90},
  number={3},
  pages={257--264},
  year={2002},
  publisher={Wiley Online Library}
}

@misc{MATLAB,
year = {2023},
author = {The MathWorks Inc.},
title = {MATLAB version: 9.14.0.2239454 (R2023a)},
publisher = {The MathWorks Inc.},
address = {Natick, Massachusetts, United States},
url = {https://www.mathworks.com}
}

@article{archontoulis2015nonlinear,
  title={Nonlinear regression models and applications in agricultural research},
  author={Archontoulis, Sotirios V and Miguez, Fernando E},
  journal={Agronomy Journal},
  volume={107},
  number={2},
  pages={786--798},
  year={2015},
  publisher={Wiley Online Library}
}
\bibliographystyle{apsrev4-2}

\end{document}